\documentclass[11pt]{article}
\usepackage{latexsym} \usepackage{epsf}
\usepackage{a4}
\usepackage{amsfonts}
\usepackage{amsmath}
\usepackage{amsthm}
\usepackage{amssymb}
\usepackage{graphics}
\usepackage{cite}

\renewcommand{\theequation}{\thesection.\arabic{equation}}
\newcounter{subequation}[equation]
\makeatletter

\expandafter\let\expandafter\reset@font\csname reset@font\endcsname

\def\subeqnarray{\arraycolsep1pt
    \def\@eqnnum\stepcounter##1{\stepcounter{subequation}%
        {\reset@font\rm(\theequation\alph{subequation})}}
\jot5mm     \eqnarray}

\makeatother

\def\tr{\mathop{\hbox{\rm tr}}\nolimits}

\def\be{\begin{equation}}
\def\ee{\end{equation}}
\def\bea{\begin{eqnarray}}
\def\eea{\end{eqnarray}}
\def\dd{\partial}
\def\half{\frac{1}{2}}

\def\one#1{#1^{\raise5pt\hbox{$\scriptstyle\!\!\!\!1$}}\,{}}
\def\two#1{#1^{\raise5pt\hbox{$\scriptstyle\!\!\!\!2$}}\,{}}

\def\II{\hbox{{1}\kern-.25em\hbox{l}}}
\makeatletter
\def\binrel@#1{\begingroup
  \setboxz@h{\thinmuskip0mu
    \medmuskip\m@ne mu\thickmuskip\@ne mu
    \setbox\tw@\hbox{$#1\m@th$}\kern-\wd\tw@
    ${}#1{}\m@th$}%
  \edef\@tempa{\endgroup\let\noexpand\binrel@@
    \ifdim\wdz@<\z@ \mathbin
    \else\ifdim\wdz@>\z@ \mathrel
    \else \relax\fi\fi}%
  \@tempa
}
\let\binrel@@\relax
\def\overset#1#2{\binrel@{#2}%
  \binrel@@{\mathop{\kern\z@#2}\limits^{#1}}}
\def\underset#1#2{\binrel@{#2}%
  \binrel@@{\mathop{\kern\z@#2}\limits_{#1}}}
\makeatother
\newfont{\bbd}{msbm10 scaled\magstep1}
\def\C{\hbox{\bbd C}}

\def\R{\hbox{\bbd R}}

\def\S{\hbox{\bbd S}}

\begin{document}

\begin{titlepage}

\vspace*{1cm}

\begin{center}
{\LARGE \bf{ Baxter operators for arbitrary  spin }}

\vspace{1cm}

{\large \sf D. Chicherin$^{da}$\footnote{\sc e-mail:chicherin@pdmi.ras.ru},
  S. Derkachov$^{a}$\footnote{\sc e-mail:derkach@pdmi.ras.ru}, D.
Karakhanyan$^b$\footnote{\sc e-mail: karakhan@lx2.yerphi.am},
R. Kirschner$^c$\footnote{\sc e-mail:Roland.Kirschner@itp.uni-leipzig.de} \\
}

\vspace{0.5cm}

\begin{itemize}
\item[$^a$]
{\it St. Petersburg Department of Steklov Mathematical Institute
of Russian Academy of Sciences,
Fontanka 27, 191023 St. Petersburg, Russia}
\item[$^b$]
{\it Yerevan Physics Institute, \\
Br. Alikhanian st. 2, 375036 Yerevan, Armenia}
\item[$^c$]
{\it Institut f\"ur Theoretische
Physik, Universit\"at Leipzig, \\
PF 100 920, D-04009 Leipzig, Germany}
\item[$^d$]
{\it Chebyshev Laboratory, St.-Petersburg State University,\\
14th Line, 29b, Saint-Petersburg, 199178 Russia}
\end{itemize}
\end{center}
\vspace{0.5cm}
\begin{abstract}
We construct Baxter operators for the homogeneous closed
$\mathrm{XXX}$ spin chain with the quantum space carrying infinite
or finite dimensional $s\ell_2$ representations. All algebraic
relations of Baxter operators and transfer matrices are deduced
uniformly from Yang-Baxter relations of the local building blocks
of these operators. This results in a systematic and very
transparent approach where the cases of finite and infinite
dimensional representations are treated in analogy. Simple
relations between the Baxter operators of both cases are obtained.
We represent the quantum spaces by polynomials and build the
operators from elementary differentiation and multiplication
operators. We present compact explicit formulae for the action of
Baxter operators on polynomials.
\end{abstract}

\vspace{4cm}

\end{titlepage}

\newpage

{\small \tableofcontents}
\renewcommand{\refname}{References.}
\renewcommand{\thefootnote}{\arabic{footnote}}
\setcounter{footnote}{0} \setcounter{equation}{0}

\section{Introduction }
\setcounter{equation}{0}

The quantum inverse scattering method
(QISM)~\cite{FST,TTF,KuSk1,Fad} is the modern approach to the
theory of integrable systems.  In the framework of QISM,
eigenstates $|v_1,...,v_k\rangle$ of the set of commuting
operators are obtained by the algebraic Bethe ansatz~(ABA) method
as  excitations over a formal vacuum state and the spectral
problem is reduced to the set of algebraic Bethe equations for the
parameters $v_j$.

The basic tools for analyzing a quantum spin chain are the
monodromy matrix $\mathbb{T}(u)$ constructed as the product of the
$\mathrm{L}$-operators referring to the sites of the chain
\begin{equation}
\label{monodr} \mathbb{T}(u)\equiv \mathrm{L}_{1}(u)
\mathrm{L}_{2}(u)... \mathrm{L}_{n}(u) = \left (\begin{array}{cc}
\mathrm{A}(u) & \mathrm{B}(u) \\
\mathrm{C}(u) & \mathrm{D}(u) \end{array} \right )
\end{equation}
and the transfer matrix $\mathrm{t}(u)$ involving
the family of commuting operators,  $\mathrm{t}(u)\,\mathrm{t}(v) =
\mathrm{t}(v)\, \mathrm{t}(u) $ constructed as the matrix trace
\begin{equation}
\label{trans} \mathrm{t}(u)\equiv \tr \mathbb{T}(u) =
\mathrm{A}(u) + \mathrm{D}(u)\,.
\end{equation}
Usually there exists some reference state $|0\rangle$ playing the
role of lowest weight vector:
$$
\mathrm{B}(u)\,|0\rangle = 0 \ ;\  \mathrm{A}(u)\,|0\rangle=
\Delta_{+}(u)\,|0\rangle \ ;\ \mathrm{D}(u)\,|0\rangle=
\Delta_{-}(u)\,|0\rangle\,,
$$
where $\Delta_{\pm}(u)$  are functions of spectral
parameter $u$.

In the ABA approach one shows that the vector $\,
|v_1,...,v_k\rangle = {\rm C}(v_1)\cdots{\rm C}(v_k)\,|0\rangle$
is an eigenvector of the operator $\mathrm{t}(u)$ with the
eigenvalue $\tau_k(u)$:
\begin{equation}
\label{1} \mathrm{t}(u) |v_1,...,v_k\rangle = \tau_k(u)
|v_1,...,v_k\rangle\ \ ;\ \ \tau_k(u)= \Delta_{+}(u)
\,\frac{Q^{(k)} (u+1)}{Q^{(k)} (u)} +
\Delta_{-}(u)\,\frac{Q^{(k)}(u-1)}{Q^{(k)}(u)}
\end{equation}
if the parameters $v_i$ obey the Bethe equations:
\begin{equation}
\label{2} Q^{(k)} (v_i+1)\,\Delta_{+}(v_i) +
Q^{(k)} (v_i-1)\,\Delta_{-}(v_i) = 0\,.
\end{equation}
All information about parameters $v_i$ is accumulated in a
polynomial $Q^{(k)} (u)=(u-v_1)\cdots(u-v_k)$.

One can obtain the Bethe equation from the formula for $\tau(u)$ by
taking residue at $u=v_i$ and using the fact that the polynomial
$\tau(u)$ is regular at this point. Finally we see that
equations~(\ref{1}) and~(\ref{2}) are equivalent to the Baxter
equation for the polynomial $Q^{(k)} (u)$,
\begin{equation}
\label{Baxter} \tau_k(u)\,Q^{(k)}(u)=\Delta_{+}(u)\,Q^{(k)} (u+1) +
\Delta_{-}(u)\,Q^{(k)} (u-1).
\end{equation}
There exists an alternative approach to the solution of the model
-- the method of $\mathrm{Q}$-operators.

In this approach
the whole problem is reduced to the construction of the Baxter
$\mathrm{Q}$-operator. In general it is an  operator $\mathrm{Q}(u)$ with the
properties~\cite{Baxter}:
\begin{itemize}
\item commutativity
$$ [ \mathrm{Q}(u) , \mathrm{Q}(v)] = 0  \ ; \
[\mathrm{Q}(u), \mathrm{t}(v)] = 0
$$
\item  finite-difference Baxter equation
\begin{equation}\label{Baxter equation}
\mathrm{t}(u)\cdot\mathrm{Q}(u) = \Delta_{+}(u) \mathrm{Q}(u+1) +
\Delta_{-}(u)\mathrm{Q}(u-1)\,.
\end{equation}
\end{itemize}
Note that in this approach  the meaning of the function
$Q^{(k)} (u)=(u-v_1)\cdots(u-v_k)$  is the polynomial eigenvalue of
the Q-operator
\begin{equation}\label{Qk}
\mathrm{Q}(u)\,|v_1,...,v_k\rangle = Q^{(k)}(u)\,|v_1,...,v_k\rangle
\,.
\end{equation}
The concept of $\mathrm{Q}$-operators has been introduced by
Baxter analyzing the eight-vertex model \cite{Baxter}.

Such operators have been studied in a number of particular
models\cite{Baxter:1972,BzSt90,GP92,Volkov,BLZ,SDQ,
KSS,Pronko,Zab,Backlund,RW,KMS,KiMa,Kor,ByTe},
where we quote here only some papers where  models with the
simplest symmetry group of the rank one are considered.

Using a $\mathrm{Q}$-operator one can perform the transformation
to the separated variable representation
\cite{Sklyanin,Skl1,DKMI}, where the eigenfunctions appear in
factorized form.

In spite of such a variety of results and long history there is so
far no commonly accepted  general scheme of $\mathrm{Q}$-operator
construction. Also the relation between different approaches to
the problem has not been discussed exhaustively.

In this paper we pursue two purposes.

At first we carry out a systematization of the results of
\cite{Derkachov:2005,DeMa,DKK} where the case of infinite
dimensional representations of the symmetry group with rank one
has been considered.

We simplify that construction significantly proving all defining
properties of $\mathrm{Q}$-operators uniformly. Namely we derive
each global relation (involving the chain operators,
$\mathrm{Q}$-operators , transfer matrices) from a corresponding
local relation for their building blocks ($\mathrm{R}$-operators
and $\mathrm{L}$-operators). In other words we pull   down
the whole construction systematically to the local level such that
properties of the global objects become more transparent. Secondly
using this construction we consider finite dimensional
representation in the quantum space and obtain corresponding
$\mathrm{Q}$-operators. That is we propose a solution of the
problem for any integer or half-integer spin as well. It is
important that the $\mathrm{Q}$-operators for representations of
both types are connected intrinsically in our approach. We
restrict ourself to the $\mathrm{XXX}$ homogeneous closed spin
chain in this paper in order to make our argumentation clearer.

In~\cite{Shortcut} Baxter operators
have been considered for the closed spin-$\half$
chain. We decided to devote a separate
paper~\cite{II} to the comparison of our approach to the ones
developed there.

The plan of the paper is the following. The first part is devoted
to a generic situation: the spin parameter $\ell$ is an arbitrary
complex number and representations are infinite-dimensional.

In a first step we construct all needed local objects --
the $\mathrm{L}$-operators and the general $\mathrm{R}$-operators.
The important tool on this stage is the construction of the
$\mathrm{R}$-operator in a factorized form. In our formulation
operators are written in terms of canonical pairs $z, \dd$ :
$[\,\dd\,, z\,] = 1$ and representation spaces are spanned by
monomials in $z$. This should not lead to misunderstandings
compared to some other papers where the notations $\hat a, \hat
a^{\dagger}$ are preferred instead.

Next we distract for a while from the systematic exposition and
demonstrate  the factorization of $\mathrm{R}$-operators at work. We show
that all needed ingredients for the construction of a
$\mathrm{Q}$-operator are present on this stage already: we
construct a $\mathrm{Q}$-operator and derive a useful formula for
its action on the generating function of monomials.
This construction of
$\mathrm{Q}$-operator does not explain its origin and has some
disadvantages from the technical point of view because the proof
of commutativity is not so simple.

So we return to  the general line and proceed to the construction
of the global objects of the chain. The general transfer matrix
$\mathrm{T}_{s} (u)$ is obtained by replacing in the transfer
matrix expression at each site $k$ the $\mathrm{L}$-matrix by the
general Yang-Baxter $\mathrm{R}$-operator acting on the tensor
product of representation modules with spins $\ell , s$ and taking
trace in the representation space $s$. Like the ordinary transfer
matrix all these operators are generating functions of the set of
commuting operators of the quantum spin chain. It turns out that
the global objects inherit factorization properties from the local
objects: the general transfer matrix $\mathrm{T}_{s}(u)$ is
factorized into a product of  simpler operators, the
$\mathrm{Q}$-operators. At this moment the $\mathrm{Q}$-operators
are assigned their natural place in a general picture.

From this point of view the general transfer matrix
$\mathrm{T}_{s}(u)$ is a one-parametric set of Baxter
$\mathrm{Q}$-operators. It is convenient to construct Baxter
operators of simpler structure which can be understood as
particular cases or restrictions of the former.

In the second part of the paper using this construction we
consider finite dimensional representation in the quantum space
and obtain corresponding $\mathrm{Q}$-operators. That is we
propose the construction for any integer or half-integer spin as well.
We formulate explicitly the intrinsic connection
between the  $\mathrm{Q}$-operators for representations of
both types.

According to our ideology we start from the general
$\mathrm{R}$-operator and restrict it to finite-dimensional
invariant subspace. The restricted $\mathrm{R}$-operators are our
local building blocks for construction of appropriate general
transfer matrices and Baxter operators in the cases of integer or
half-integer spin representation in the quantum space. The
derivation of factorization of the general transfer matrix into
the product of $\mathrm{Q}$-operators follows step by step the
corresponding calculation in the generic spin case. The same is
true for the other properties of Baxter operators. Then we
establish the connections between compact spin $\mathrm{Q}$-operators
and limits of Baxter operators for generic spin. We show that a
 careful  calculation of limits with $\mathrm{Q}$-operators
for infinite-dimensional representations results in the
appropriate $\mathrm{Q}$-operators for finite-dimensional
representations. This leads us to fairly simple, compact formulae
for the Baxter operators for  integer or half-integer spin.

\newpage

\section{Local objects: quantum $\mathrm{L}$-operator
and general $\mathrm{R}$-operator}

\setcounter{equation}{0}

We consider first the operators representing the local building
units of the chains. Actually the $\mathrm{L}$-operator contains
the local information about the system and the $\R$-operators are
derived therefrom. On the other hand  the $\R$-operators are more
convenient as building units acting on the tensor product of
quantum and auxiliary spaces carrying arbitrary representations.
The particular case of  spin $\half$ representation in one of
these spaces leads us back to the $\mathrm{L}$-matrix. The
$\R$-operators provide  us the starting point from which the
different versions of Baxter operators can be obtained.

\subsection{$\mathrm{L}$-operators}
\label{L-operators}

We consider spin chains with $s\ell_2$ symmetry algebra and use
the functional representation of the algebra in the space of
polynomials of one complex variable $\C[z]$. Fixing  the generic
complex number $\ell$ and representing generators of the algebra
as differential operators of the first order \be
\label{generators} \mathrm{S} = z\dd -\ell\ ,\ \mathrm{S}_{-} =
-\dd \ ,\ \mathrm{S}_{+} = z^2\dd - 2\ell z \ee we provide $\C[z]$
with the structure of a Verma module with lowest weight $-\ell$ which we
denote in the following by $\mathbb{U}_{-\ell}$. We keep this
unusual sign for remembering that we are working with lowest
weight vectors instead of highest weight vectors.

The module $\mathbb{U}_{-\ell}$ is the infinite-dimensional
functional space with the basis $\{1, z , z^2 , z^3 \cdots\}$ and
there is no invariant subspace in the module for generic $\ell$,
in other words it is irreducible. An invariant finite-dimensional
subspace $\mathbb{V}_n$ appears for special values $\ell =
\frac{n}{2}, n = 0,1,2,3\cdots$. It is the $(n+1)$-dimensional
irreducible submodule with the basis $\{1, z,\cdots z^{n}\}$. The
infinite-dimensional quotient module $\mathbb{U}_{\frac{n}{2}+1} =
\mathbb{U}_{-\frac{n}{2}}/\mathbb{V}_n$ with basis $\{z^{n+1} ,
z^{n+2} , \cdots \}$ is also irreducible. Its lowest weight is
$\frac{n}{2}+1$.

Representing generators as differential operators is very
convenient since it allows to describe finite-dimensional ({\it
compact spin}) and infinite-dimensional ({\it non-compact spin})
representations at once. Following this idea  further will allow
us to construct $\mathrm{Q}$-operators for both types of
representations in the quantum space in a compact and clear way.

We recover the $s\ell_2$-generators $\mathbf{s} ,
\mathbf{s}_{\pm}$ in fundamental representation from expressions
for generators in generic representations~(\ref{generators}) for
$\ell = \frac{1}{2}$ with the basis $ \mathbf{e}_1 =
\mathrm{S}_{+}\cdot 1 = - z \ ,\ \mathbf{e}_2 = 1 $ in the known
Pauli-matrix form
$$
\mathbf{s} = \frac{1}{2}\left(\begin{array}{cc}
1 & 0\\
0 & -1
   \end{array}\right)\ ,\ \mathbf{s}_{-} =
   \left(\begin{array}{cc}
0 & 0 \\
1 & 0
   \end{array}\right)\ ,\ \mathbf{s}_{+} =
\left(\begin{array}{cc}
0 & 1 \\
0 & 0
   \end{array}\right)\,,
$$
where we used the standard definition for the matrix of an operator: $
\mathrm{A} \mathbf{e}_i = \sum_{k} \mathbf{e}_k \mathrm{A}_{ki}. $

We define the $\mathrm{L}$-operator as
$$
\mathrm{L}(u) \equiv u \cdot \II \otimes \II +
2\cdot\mathrm{S}\otimes \mathbf{s}+\mathrm{S}_{-}\otimes
\mathbf{s}_{+}+\mathrm{S}_{+}\otimes\mathbf{s}_{-}
$$
in terms of generators of the algebra (\ref{generators}). It acts
in $\mathbb{U}_{-\ell} \otimes \C^2$ and depends on two
parameters: spin $\ell$ and spectral parameter $u$. It respects
Yang-Baxter equation in the space $\C^2 \otimes \C^2 \otimes
\mathbb{U}_{-\ell}$ for Yang $\mathrm{R}$-matrix \be
\mathrm{R}_{ij,nm}(u-v) \cdot \mathrm{L}_{ns}(u)\cdot
\mathrm{L}_{mp}(v) = \mathrm{L}_{is}(v)\cdot
\mathrm{L}_{jp}(u)\cdot \mathrm{R}_{sp,nm}(u-v) \label{FCR} \ee
where $i,j, \cdots = 1,2$ and $ \mathrm{R}_{ij,nm}(u) = u
\cdot\delta_{in}\,\delta_{jm}+\delta_{im}\,\delta_{jn}. $ This
equation also referred to as the fundamental commutation relation
contains in a compact form all  relations of underlying Yangian
symmetry algebra. Taking into account the expressions for the
generators the $\mathrm{L}$-operator can be written as a matrix
$2 \times 2$ with operational elements acting in the space
$\mathbb{U}_{-\ell}$
\begin{equation}
\mathrm{L}(u) = \left(\begin{array}{cc}
u-\ell+z\dd & -\dd \\
z^2\dd - 2\ell z& u+\ell-z\dd
   \end{array}\right)\,.
\label{Lax}
\end{equation}
There exists the useful factorized representation
\begin{equation}
\mathrm{L}(u_{1},u_{2}) = \left(%
\begin{array}{cc}
  1 & 0 \\
  z & 1 \\
\end{array}%
\right)\left(%
\begin{array}{cc}
  u_1 & -\partial \\
  0 & u_2 \\
\end{array}%
\right)\left(%
\begin{array}{cc}
  1 & 0 \\
  -z & 1 \\
\end{array}%
\right)\, . \label{LaxFact}
\end{equation}
We have introduced the parameters $u_1$ and $u_2$: $u_{1}\equiv
u-\ell-1 \ ,u_{2}\equiv u+\ell$  instead of $u$ and $\ell$ because
they are very convenient for our purposes.

\subsection{General R-operator}

In order to avoid misunderstandings we distinguish two versions of
the general Yang-Baxter operators acting in the space
$\mathbb{U}_{-\ell_1}\otimes\mathbb{U}_{-\ell_2}$ by the notations
$\mathrm{R}_{12}$ and $\R_{12}$.  The former does not contain the
permutation operator $\mathrm{P}_{12}\,$:
$\,\mathrm{P}_{12}\,\psi(z_1,z_2) = \psi(z_2,z_1)$ whereas the
latter does, so they are related by
$$
\R_{12} = \mathrm{P}_{12}\, \mathrm{R}_{12}.
$$
In this section we use only $\mathrm{R}$-operator notations
however in the subsequent sections we will see that $\R$-operator
notations are very natural for certain purposes.

The general $\mathrm{R}$-operator acts on the space
$\mathbb{U}_{-\ell_1}\otimes\mathbb{U}_{-\ell_2}$ and is defined
as the $s\ell_2$-symmetric solution of the following relation
({\it $\mathrm{RLL}$-relation})
\begin{equation}\label{RLL}
\mathrm{R}_{12}(u_{1},u_2|v_{1},v_2)\,
\mathrm{L}_{1}(u_1,u_2)\,\mathrm{L}_{2}(v_1,v_2)=
\mathrm{L}_{1}(v_1,v_2)\,\mathrm{L}_{2}(u_1,u_2)\,
\mathrm{R}_{12}(u_{1},u_2|v_{1},v_2)
\end{equation}
where \be \label{param} u_1 = u-\ell_1-1 ,\ u_2 = u+\ell_1 ;\ v_1
= v-\ell_2-1 ,\ v_2 = v+\ell_2 . \ee
The operator $\mathrm{R}$
depends on the difference of spectral parameters $u-v$ and two
spins $\ell_1,\ell_2$, therefore the defining relation can be
rewritten as
\begin{equation}
\mathrm{R}(u-v|\ell_{1},\ell_2)
\mathrm{L}_{1}(u)\,\mathrm{L}_{2}(v)=
\mathrm{L}_{1}(v)\,\mathrm{L}_{2}(u)
\mathrm{R}(u-v|\ell_{1},\ell_2).
\end{equation}
Roughly speaking the operator $\mathrm{R}$ interchanges parameters
$u_1,u_2$ with $v_1,v_2$ in the product of $\mathrm{L}$-operators.

\medskip

\medskip

\hspace{20mm}
\includegraphics{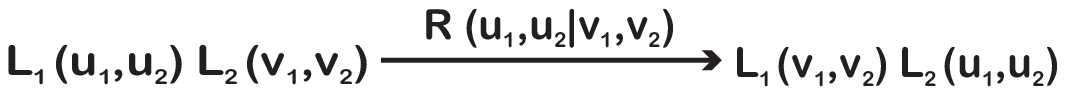}

\medskip

\medskip


It is useful to focus on this parameter exchange further and to
consider more $\mathrm{R}$-operators related to other exchange
operations on the set $(u_1,u_2,v_1,v_2)$.
 In this set of
operators there are those which interchange adjacent parameters.
These are the most elementary ones because they are the building blocks
for all other operators of parameter permutations. Thus the
complex problem of solving the general Yang-Baxter equation
reduces to a set  of  simpler ones. This idea has been carried out
in the paper~\cite{DKK}. For our current purposes we do not need
the  most elementary operators of permutations and we work with
some composite operators constructed from these elementary ones.
The main reason for this is that the elementary operators are not
well defined in the space $\C[z]$ and one would have to consider a
larger space, whereas the composite operators which we are going
to work with here are well defined in the space $\C[z]$. In order
make the previous statements explicit we collect here some useful
formulae.

Let us define the power of the derivative operator for any
$\alpha$ by
\begin{equation}\label{d}
\partial^\alpha \equiv
\frac{1}{z^\alpha}\frac{\Gamma(z\partial+1)}{\Gamma(z\partial+1-\alpha)}.
\end{equation}
Certainly, if $\alpha$ takes values $1,2,\ldots$ we obtain the
familiar  multiple derivative.

All expected properties of this operator can be easily proven
using this definition.
\begin{itemize}
    \item \it{commutativity and group property}: $\ \partial^\alpha\,\partial^\beta =
    \partial^\beta\, \partial^\alpha =
    \partial^{\alpha+\beta}\ \ ;\ \ \partial^0 = \II$
    \item {\it differentiation rule}
$\partial^\alpha\, z = z\, \partial^\alpha +\alpha
\partial^{\alpha-1}$
    \item {\it star-triangle
    relation} : $\ \partial^\alpha\, z^{\alpha+\beta}\partial^\beta =
    z^{\beta}\partial^{\alpha+\beta}\,z^\alpha$
    \item {\it connection with $\Gamma$-functions:} $z^{\beta}\partial^{\alpha+\beta}\,z^\alpha =
    \frac{\Gamma(z\partial+1+\alpha)}{\Gamma(z\partial+1-\beta)}
$
\end{itemize}
We define the operators $\mathrm{R}^{1}_{12}$ and
$\mathrm{R}^{2}_{12}$ acting in the space $\mathbb{U}_{-\ell_1}
\otimes \mathbb{U}_{-\ell_2}$  by the following relations
\begin{equation}
\mathrm{R}^{1}_{12}\,
\mathrm{L}_{1}(u_1,u_2)\mathrm{L}_{2}(v_1,v_2)=
\mathrm{L}_{1}(v_1,u_2)\mathrm{L}_{2}(u_1,v_2)\,
\mathrm{R}^{1}_{12} \label{R1}
\end{equation}
\begin{equation}
\mathrm{R}^{2}_{12}\,
\mathrm{L}_{1}(u_1,u_2)\mathrm{L}_{2}(v_1,v_2)=
\mathrm{L}_{1}(u_1,v_2)\mathrm{L}_{2}(v_1,u_2)\,
\mathrm{R}^{2}_{12} \label{R2}
\end{equation}
To avoid misunderstanding we notice that upper indices $1,2$
distinguish our two operators and lower indices  usually show
in which spaces the operators act nontrivially. In the generic situation
$\ell_1,\ell_2\in \mathbb{C}$ the space $\mathbb{U}_{-\ell_1}
\otimes \mathbb{U}_{-\ell_2}$ is isomorphic to the space
$\mathbb{C}[z_1,z_2]$ of polynomials of two variables $z_1$ and
$z_2$ and the solutions of these equations have the form
$$
\mathrm{R}^1_{12}(u_1|v_1,v_2) =
z_{21}^{v_2-v_1}\,\partial_{2}^{u_1-v_1}\, z_{21}^{u_1-v_2} =
\frac{\Gamma(z_{21}\dd_2+u_1-v_2+1)}{\Gamma(z_{21}\dd_2+v_1-v_2+1)}
$$
\be \label{R1R2} \mathrm{R}^2_{12}(u_1,u_2|v_2) =
z_{12}^{u_2-u_1}\,\partial_{1}^{u_2-v_2}\, z_{12}^{u_1-v_2} =
\frac{\Gamma(z_{12}\dd_1+u_1-v_2+1)}{\Gamma(z_{12}\dd_1+u_1-u_2+1)}
. \ee
It is evident that the operator $\mathrm{R}^{k}$ commutes with
$z_k$: $\mathrm{R}^{1}\, z_1 = z_1\, \mathrm{R}^{1}\ ;\
\mathrm{R}^{2}\, z_2 = z_2\, \mathrm{R}^{2}$.

There are  two ways to interchange parameters $(u_1,u_2,
v_1,v_2) \to (v_1,v_2,u_1,u_2)$ in the product of two
$\mathrm{L}$-operators by interchanging the ordering in the pairs
$u_1, v_1$ and $u_2,v_2$. Correspondingly the
$\mathrm{R}$-operator can be factorized in two ways as follows
\begin{equation}
\mathrm{R}(u_{1},u_2|v_{1},v_2)=
\mathrm{R}^{1}(u_1|v_1,u_2)\mathrm{R}^{2}(u_1,u_2|v_{2}) =
\mathrm{R}^{2}(v_1,u_2|v_2)\mathrm{R}^{1}(u_1|v_1,v_{2})
\label{Rfact}
\end{equation}
This can be checked with the star-triangle relation.

Now we would like to illustrate the factorization of the general
$\mathrm{R}$-operator by simple and transparent pictures. The
operator $\mathrm{R}^{1}$ interchanges $u_1$ and $v_1$:

\medskip

\medskip

\hspace{20mm}
\includegraphics{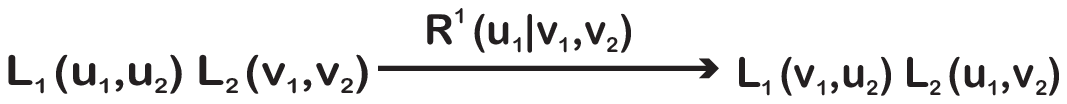}

\medskip

\medskip


and the operator $\mathrm{R}^{2}$ interchanges $u_2$ and $v_2$:

\medskip

\medskip

\hspace{20mm}
\includegraphics{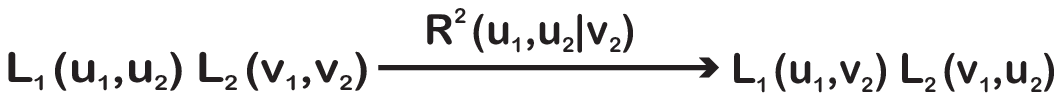}

\medskip

\medskip


The operator $\mathrm{R}^{1}\mathrm{R}^{2}$ interchanges parameters
$u_1,v_1$ and $u_2,v_2$ in two steps so that the factorization
indicated above is the condition of commutativity for the diagram

\medskip

\medskip

\hspace{20mm}
\includegraphics{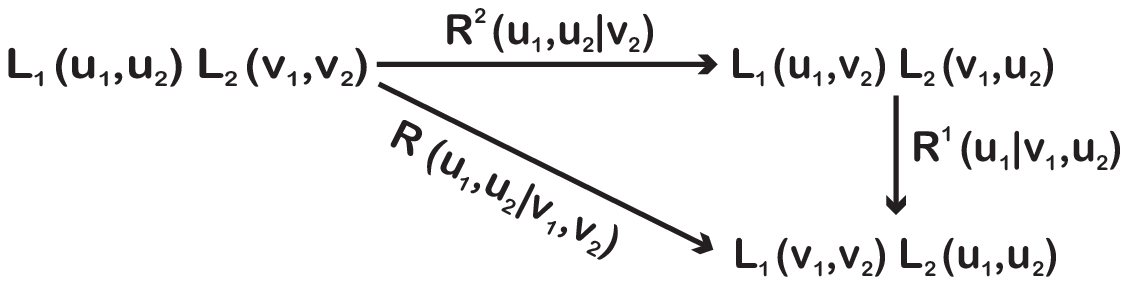}

\medskip

\medskip


Two equivalent ways to interchange parameters can be depicted by
commutative diagram

\medskip

\medskip

\hspace{10mm}
\includegraphics{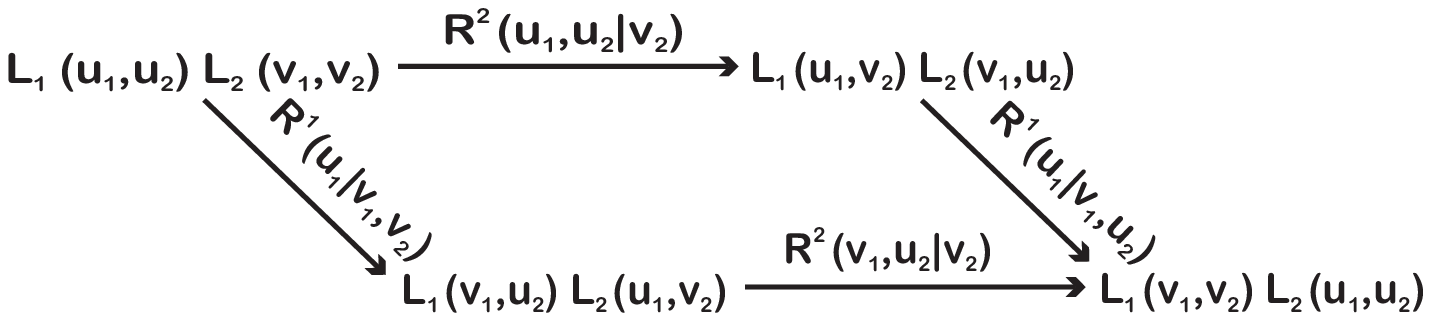}

\medskip

\medskip


which present two equivalent expressions for $\mathrm{R}$-operator
(\ref{Rfact}).

At certain parameter values the operators become simpler. Indeed,
when $u_1 = v_1$ or $u_2 = v_2$ the corresponding permutation is
trivial, consequently \be \label{trivial}
\mathrm{R}^1(u_1|u_1,v_2) = \II \, , \qquad
\mathrm{R}^2(u_1,u_2|u_2) = \II \ee and
$$
\mathrm{R}(u_1 , u_2| u_1 , v_2) = \mathrm{R}^{2}(u_1,u_2|v_{2})
\, , \qquad \mathrm{R}(u_1 , u_2| v_1 , u_2) =
\mathrm{R}^{1}(u_1|v_1,u_2).
$$

Notice that the latter relations hold for  generic parameters only
and modify in the case of integer values of $u_2-u_1$ or
$v_2-v_1$.

It is well known that along with the fundamental Yang-Baxter
relation (\ref{FCR}) and the RLL-relation (\ref{RLL}) there is
a further relation involving three general operators $\mathrm{R}$. It
can be understood as the formulation of the equivalence of two
ways of transforming $\mathrm{L}_1 (u_1,u_2) \mathrm{L}_2(v_1,v_2)
\mathrm{L}_3(w_1,w_2)$ $ \to \mathrm{L}_1 (w_1,w_2)
\mathrm{L}_2(v_1,v_2) \mathrm{L}_3(u_1,u_2)$

\begin{equation}
\label{RRR}
\mathrm{R}_{12}(v_{1},v_2|w_{1},w_2)\mathrm{R}_{23}(u_{1},u_2|w_1,w_2)
\mathrm{R}_{12}(u_1,u_2|v_{1},v_2) =
\end{equation}
$$
=\mathrm{R}_{23}(u_{1},u_{2}|v_1,v_2)\mathrm{R}_{12}(u_1,u_{2}|w_1,w_{2})
\mathrm{R}_{23}(v_{1},v_2|w_{1},w_2)
$$
This relation  can be represented by the commutative diagram

\medskip

\medskip

\includegraphics{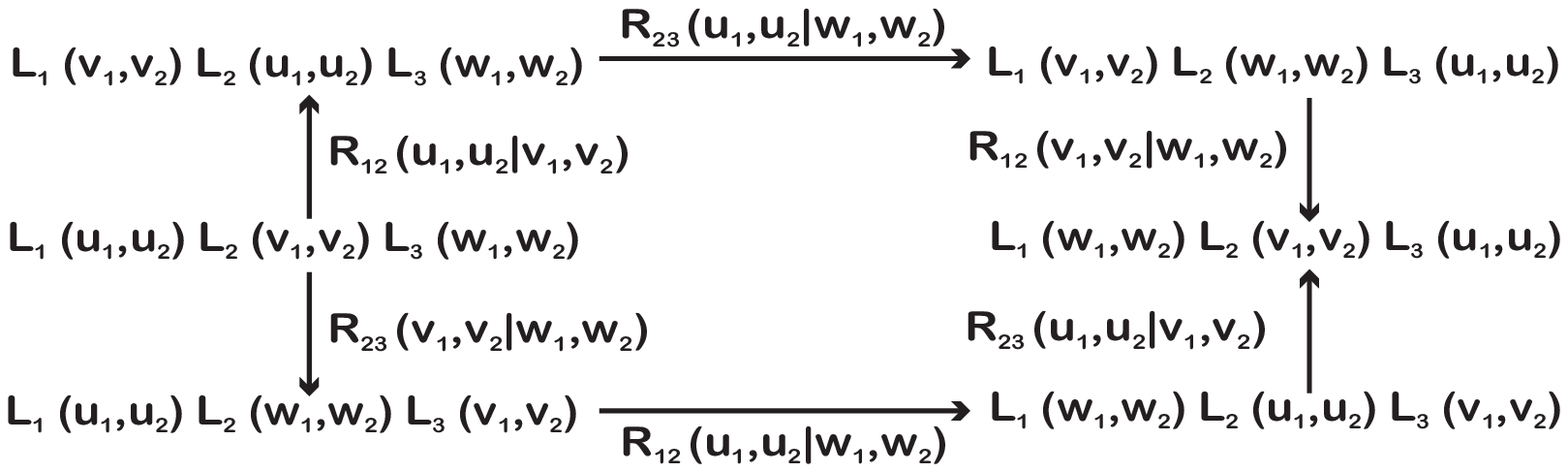}

\medskip

\medskip

A further relation involves two $\mathrm{R}$-operators and one
operator $\mathrm{R}^{2}$. In analogy it can be understood as the
formulation of the equivalence of the two ways transforming
$\mathrm{L}_1 (u_1,u_2) \mathrm{L}_2(v_1,v_2)
\mathrm{L}_3(w_1,w_2) $ $ \to \mathrm{L}_1 (v_1,w_2)
\mathrm{L}_2(w_1,v_2) \mathrm{L}_3(u_1,u_2)$
\be \label{R2RR}
\mathrm{R}^2_{12}(v_1,v_2|w_2)\mathrm{R}_{23}(u_{1},u_2|w_1,w_2)
\mathrm{R}_{12}(u_{1},u_2|v_1,v_{2}) =  \ee
$$
=\mathrm{R}_{23}(u_{1},u_2|w_1,v_{2})\mathrm{R}_{12}(u_1,u_{2}|v_1,w_{2})
\mathrm{R}^2_{23}(v_{1},v_{2}|w_2)
$$

depicted by the diagram

\medskip

\medskip

\includegraphics{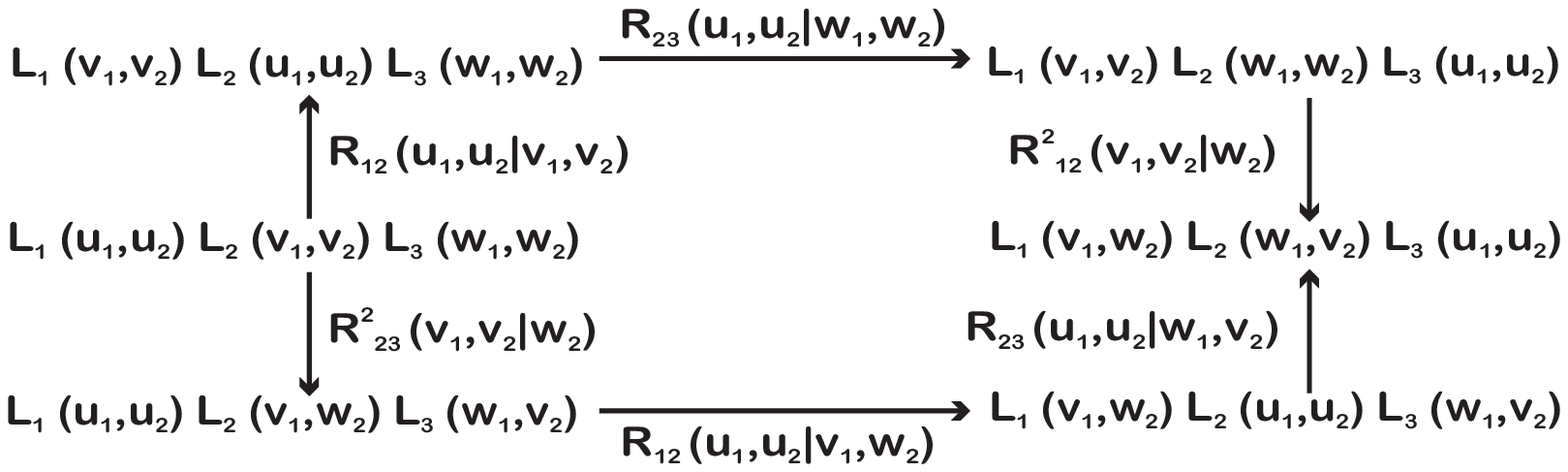}

\medskip

\medskip

We have a third relation where instead of $\mathrm{R}^2$ the other
factor operator $\mathrm{R}^1$ is involved. \be \label{R1RR}
\mathrm{R}^1_{12}(v_1|w_1,w_2)\mathrm{R}_{23}(u_{1},u_2|w_1,w_2)
\mathrm{R}_{12}(u_{1},u_2|v_1,v_{2}) =  \ee
$$
=\mathrm{R}_{23}(u_{1},u_2|v_1,w_{2})\mathrm{R}_{12}(u_1,u_{2}|w_1,v_{2})
\mathrm{R}^1_{23}(v_{1}|w_{1},w_2)\, .
$$
All these relations can be proved directly by using explicit
expressions for all operators~(\ref{R1R2},\ref{Rfact}) but, as
V.Tarasov explained  us, the very existence of operators $\R^1$
and $\R^2$ and all relations among them can be extracted from the
general theory developed in papers~\cite{Tar85}.

Note that the general Yang-Baxter equation~(\ref{RRR}) is the
consequence of~(\ref{R2RR}) and~(\ref{R1RR}) but the two latter
are not a straightforward  consequence of the general Yang-Baxter
equation. Furthermore for our purpose they are as
 important as the general Yang-Baxter equation.
Indeed, we will show that starting just only from these three
relations (\ref{RRR}),(\ref{R2RR}),(\ref{R1RR}) it is possible to
deduce factorization of the general transfer matrix
$\mathrm{T}(u)$ into the product of Baxter $\mathrm{Q}$-operators,
commutativity of all these operators and also to obtain the Baxter
relation.
 No further constructions or concepts are needed.


\section{A $\mathrm{Q}$-operator constructed from $\mathrm{R}^2$}

\label{Qbyhand}

\setcounter{equation}{0}

In this section we would like to show the introduced operators at
work. For this purpose we construct a $\mathrm{Q}$-operator using
the operators $\mathrm{R}^2$ as building blocks. Later we shall
present a more systematic and exhaustive construction.  But now in
this simple example we would like to show that the properties of a
$\mathrm{Q}$-operator follow directly from properties of its local
building blocks.

\subsection{Transfer matrix $\mathrm{t}(u)$}

\label{tm}

The closed homogeneous $s\ell_2$-symmetric spin chain under
consideration consists of $N$ sites carrying representations of
 the same representation parameter $\ell$.
The operator $\mathrm{L}_k(u)$ (\ref{Lax}) acts in
the space $\mathbb{U}_{-\ell} \otimes \C^2$. The corresponding
quantum space $\mathbb{U}_{-\ell}$ is associated with each site
and for generic $\ell\in \mathbb{C}$ is isomorphic to the space
$\mathbb{C}[z_k]$ of polynomials of the variable $z_k$. The space
$\mathbb{C}^2$ is a common auxiliary space. Then the product
$\mathrm{L}_1(u)\cdots \mathrm{L}_N(u)$ acts in the space
$\mathbb{U}_{-\ell} \otimes \ldots \otimes\mathbb{U}_{-\ell}
\otimes \C^2$. Taking trace over the auxiliary two-dimensional space
$\C^2$ we define the transfer matrix $\mathrm{t}(u)$
\begin{equation}
\label{tmat} \mathrm{t}(u) = \tr
\mathrm{L}_1(u)\mathrm{L}_2(u)\cdots \mathrm{L}_N(u)
\end{equation}
Recall the standard argument for commutativity of the ordinary
transfer matrix, \be [\mathrm{t}(u), \mathrm{t}(u)] = 0 \ee The
fundamental Yang-Baxter relation (\ref{FCR}) implies the analogous
relation with the $\mathrm{L}$ matrices replaced by the monodromy
matrix, i.e. the product of $N$ matrices $\mathrm{L}_k$, where the
operator matrix elements act in $\mathbb{U}_{-\ell}\sim
\mathbb{C}[z_k]$, \be \label{tstand}
 \mathrm{L}_{ij} \rightarrow \mathrm{L}_{i k_1} \mathrm{L}_{k_1 k_2}...
\mathrm{L}_{k_{N-1} j}
\ee
Then taking  traces in both 2-dimensional tensor factors results in the
vanishing commutator of the transfer matrices.
The proofs of factorization and commutativity for the other global
spin chain
operators to be defined below follow basically this scheme starting from
the appropriate form of the Yang-Baxter relation.

Since $\mathrm{t}(u)$ is polynomial in the spectral parameter we
obtain the family of $N-1$ commuting operators acting in the space
$\mathbb{U}_{-\ell} \otimes \ldots \otimes\mathbb{U}_{-\ell}$. In
the generic situation $\ell\in \mathbb{C}$ this quantum space is
isomorphic to the space $\mathbb{C}[z_1,\cdots,z_N]$ of
polynomials depending on variables $z_1,\cdots,z_N$. The reference
state $|0\rangle$ of ABA plays the common lowest weight vector for
all representations $\mathbb{U}_{-\ell}$:
$\mathrm{S}^{-}_k\,|0\rangle = 0$, i.e. the polynomial of zero
degree, the  constant function. The eigenvalues of the operators
$\mathrm{A}(u)$ and $\mathrm{D}(u)$ appearing in~(\ref{monodr})
can be easily calculated so that we have explicit expressions for
the functions $\Delta_{\pm}(u)$: $\,\Delta_{\pm}(u) =
(u\mp\ell)^N$.

\subsection{$\mathrm{Q}$-operator and Baxter equation}

The left hand side of the Baxter equation~(\ref{Baxter equation})
 involves the product of the transfer matrix $\mathrm{t}(u)$ and the
$\mathrm{Q}$-operator. The former is constructed from
$\mathrm{L}$-operators and we are going to construct the latter
from operators $\mathrm{R}^2$. Thus we need a local relation which
comprise the product of $\mathrm{R}^2$ and $\mathrm{L}$ operators. It
becomes clear soon that the defining equation for the operator
$\mathrm{R}^2$ (\ref{R2}) fits very well for this purpose.

Using the factorization formulae (\ref{LaxFact}) for
$\mathrm{L}_1(u_1,u_2)$ and $\mathrm{L}_2(v_1,v_2)$ and the
commutativity $\mathrm{R}^{2}\, z_2 = z_2\, \mathrm{R}^{2}$ we
rewrite (\ref{R2}) in a slightly different form
$$
\mathrm{Z}_1^{-1}\,
\mathrm{R}^{2}_{12}(u-v_2)\,\mathrm{L}_1(u_1,u_2)\,\mathrm{Z}_2
= \left(%
\begin{array}{cc}
  u_1 & -\partial_{1} \\
  0 & v_2 \\
\end{array}%
\right)\left(%
\begin{array}{cc}
  1 & 0 \\
  -z_{12} & 1 \\
\end{array}%
\right)\left(%
\begin{array}{cc}
  v_1 & -\partial_{2} \\
  0 & u_2 \\
\end{array}%
\right)\cdot\mathrm{R}^{2}_{12}(u-v_2)\cdot \left(%
\begin{array}{cc}
  v_1 & -\partial_{2} \\
  0 & v_2 \\
\end{array}%
\right)^{-1}\,.
$$
We see that the dependence of  $\mathrm{R}^{2}$ on
the parameter $v_2$ enters by a  shift of the spectral
parameter $u$. $\mathrm{Z}_k$ are triangular matrices.
$$
\mathrm{R}^{2}_{12}(u-v_2)=
\partial_{1}^{u_1-v_2}z_{12}^{u_2-v_2}
\partial_{1}^{u_2-u_1}\ \ ;\ \ \mathrm{Z}_k \equiv \left(%
\begin{array}{cc}
  1 & 0 \\
  z_k & 1 \\
\end{array}%
\right)\,.
$$
Next we have to calculate the product of matrices at right
hand side. For our purpose we need only the
diagonal elements of the matrix. We obtain as upper diagonal
element $( u_1 + \partial_1 z_{12}) \, \mathrm{R}^2_{12}(u-v_2)$
which is transformed to the needed form using the relation
$$
\partial_1 z_{12} \, \mathrm{R}^2_{12}(u-v_2) =
\mathrm{R}^2_{12}(u+1-v_2) - (u_1 - v_2) \,
\mathrm{R}^2_{12}(u-v_2)\,.
$$
The lower diagonal element, $ u_2 \, \mathrm{R}^2_{12}(u-v_2) +
z_{12} \, [\partial_2 \, , \mathrm{R}^2_{12}(u-v_2)] $,  is
transformed to the final form by using the relation
$$
z_{12} \, [\partial_2 \, , \mathrm{R}^2_{12}(u-v_2)] =
(v_2-u_2)\mathrm{R}^2_{12}(u-v_2) + (u_1-v_2)(u_2 - v_2)
\,\mathrm{R}^2_{12}(u-1-v_2)
$$
Performing this calculation we use only the simple commutation
properties of the operator (\ref{d}). Finally we obtain
$$
\mathrm{Z}_1^{-1}\,
\mathrm{R}^{2}_{12}(u-v_2)\,\mathrm{L}_1(u_1,u_2)\,\mathrm{Z}_2 =
$$
$$
= \left(%
\begin{array}{cc}
\mathrm{R}^{2}_{12}(u+1-v_2) +
v_2\mathrm{R}^{2}_{12}(u-v_2) & -\mathrm{R}^{2}_{12}(u-v_2)\partial_{1} \\
-v_2\,z_{12}\,\mathrm{R}^{2}_{12}(u-v_2) & (u_1-v_2)(u_2-v_2)
\mathrm{R}^{2}_{12}(u-1-v_2) +
v_2 \mathrm{R}^{2}_{12}(u-v_2) \\
\end{array}%
\right).
$$
The crucial  feature of this relation being the cornerstone of the
current calculation is that at the point $v_2=0$ the matrix on
right hand side becomes upper triangular.  We put $v_2=0$ in the
derived matrix relation and specify it by choosing the first space
to be the local quantum space at site $k$ and the second space the local
quantum space at site $(k+1)$
$$
\mathrm{Z}^{-1}_k\cdot\mathrm{R}^{2}_{k k+1}(u)
\,\mathrm{L}_k(u_1,u_2)\cdot\mathrm{Z}_{k+1} =
\left(%
\begin{array}{cc}
\mathrm{R}^{2}_{k k+1}(u+1) & -\mathrm{R}^{2}_{k k+1}(u)\partial_{k} \\
 0 & u_1\,u_2\,\mathrm{R}^{2}_{k k+1}(u-1) \\
\end{array}%
\right)\,.
$$
This is exactly the local relation which underlies Baxter equation.
We take the product over all sites
\be \label{mainf}
\mathrm{Z}_1^{-1}\cdot\mathrm{R}_{12}^{2}(u)
\mathrm{R}_{23}^{2}(u) \cdots \mathrm{R}_{N0}^{2}(u) \cdot
\mathrm{L}_1(u)\mathrm{L}_2(u)\cdots\mathrm{L}_N(u)\cdot\mathrm{Z}_0
=
\ee
$$
=
\left(%
\begin{array}{cc}
\mathrm{R}^{2}_{12}(u+1) & -\mathrm{R}^{2}_{12}(u)\partial_{1} \\
 0 & u_1\,u_2\,\mathrm{R}^{2}_{12}(u-1) \\
\end{array}%
\right)\cdots\left(%
\begin{array}{cc}
\mathrm{R}^{2}_{N0}(u+1) & -\mathrm{R}^{2}_{N0}(u)\partial_{N} \\
 0 & u_1\,u_2\, \mathrm{R}^{2}_{N0}(u-1) \\
\end{array}%
\right)
$$
appending one additional fictitious site $0$. In this product the
matrices $\mathrm{Z}_k$ and $\mathrm{Z}^{-1}_k$ ($k=2,3,\ldots,N$)
cancel pairwise. Then we calculate the trace over the
two-dimensional space $\C^2$, use commutativity of all
$\mathrm{R}^{2}$ and $\mathrm{L}_k$ with $z_0$ in order to move
$\mathrm{Z}_{0}$ to the left and finally identify sites $0$ and
$1$. Thus we obtain the Baxter equation
\begin{equation}\label{Baxter2}
\mathrm{t}(u)\mathrm{Q}_2(u) = \mathrm{Q}_2(u+1)+ (u_1 u_2)^N
\cdot\mathrm{Q}_2(u-1)
\end{equation}
for the operator\footnote{We also multiply the last equation by
cyclic shift
$\mathrm{P}=\mathrm{P}_{12}\mathrm{P}_{13}\cdots\mathrm{P}_{1N}$
for later convenience.}
\begin{equation}\label{Q_2}
\mathrm{Q}_2(u) = \mathrm{P}\cdot\left.\mathrm{R}_{12}^{2}(u)\,
\mathrm{R}_{23}^{2}(u)\, \cdots\, \mathrm{R}_{N-1,N}^{2}(u)
\mathrm{R}_{N0}^{2}(u)\right|_{z_0\to z_1}\ .
\end{equation}
We have constructed explicitly a solution of the Baxter equation
relying on the local relation (\ref{R2}) only. It is also
possible to check in this way the commutativity properties of
$\mathrm{Q}_{2}(u)$ but we would like to obtain this result  in
the framework of the general scheme
to be considered further. Note that the Baxter equation~(\ref{Baxter})
can be transformed to the canonical form by appropriate
normalization of the operator $\mathrm{Q}_{2}(u)$ which will
 appear naturally in the next subsection.

\subsection{Explicit formulae for the action of $\mathrm{Q}_{2}$ on polynomials}

Now we would like to visualize the constructed operator
$\mathrm{Q}_{2}$. In order to use the advantages of the generating
function method we combine the basis vectors of the module
$\mathbb{U}_{-\ell}$ into
\begin{equation}\label{gen}
e^{x \mathrm{S}^{+}} \cdot 1 = ( 1 - x z )^{2 \ell}
\,,
\end{equation}
where $x$ is an auxiliary parameter.  The derivative $\partial_x^k$
at the point $x=0$ produces the basis vector $\mathrm{S}_+^k\cdot 1
\sim z^k$. We shall obtain a very simple representation for
$\mathrm{Q}_{2}(u)$ acting on the global generating function $(1-x_{1}
z_{1})^{2\ell}\cdots(1-x_{N} z_{N})^{2\ell}$. This formula
contains in transparent form all information about the action of
the operator $\mathrm{Q}_2(u)$ on polynomials. Indeed, the calculation
of the derivative $\partial_{x_1}^{k_1}\cdots\partial_{x_N}^{k_N}$
at  $x_1=\ldots=x_N=0$ results in the explicit expression for
 the action of the operator $\mathrm{Q}_2(u)$ on the
monomial $z_1^{k_1}\cdots z_N^{k_N}$.

 The  special construction of the operator
$\mathrm{Q}_2(u)$ reduces the global problem to a local one.
Indeed the whole expression factorizes into the pieces of the
simple form
\begin{equation}\label{Qgenerat}
\mathrm{R}^{2}_{1 2} (u) \, \mathrm{R}^{2}_{2 3} (u) \, \cdots \,
\mathrm{R}^{2}_{N 0} (u) \cdot (1-x_{1}z_{1})^{2\ell} (1-x_{2}
z_{2})^{2\ell} \cdots(1-x_{N} z_{N})^{2\ell} =
\end{equation}
$$
= \mathrm{R}^{2}_{1 2}(u) \, (1-x_{1}z_{1})^{2\ell} \cdot
\mathrm{R}^{2}_{2 3}(u) \, (1-x_{2} z_{2})^{2\ell} \cdots
\mathrm{R}^{2}_{N 0}(u) \, (1-x_{N} z_{N})^{2\ell}
$$
so that we have to calculate the local quantity $\mathrm{R}^{2}_{k
k+1}(u) \, (1-x_{k} z_{k})^{2\ell}$. It turns out that this
expression can be obtained from  "first symmetry principles"
almost without calculations.

We start with some intertwining properties of $\mathrm{R}^2$.
Since it depends on the difference of the spectral
parameters the shift $u \to u + \lambda  \ , \, v \to v + \lambda$
does not affect $\mathrm{R}^2_{12}$. Thus performing this shift in
the defining relation (\ref{R2}) and collecting terms linear in
$\lambda$ we obtain
$$
\mathrm{R}^{2}_{12} \cdot \left( \mathrm{L}_{1}(u_1,u_2) +
\mathrm{L}_{2}(v_1,v_2)\right)= \left( \mathrm{L}_{1}(u_1,v_2) +
\mathrm{L}_{2}(v_1,u_2) \right) \cdot \mathrm{R}^{2}_{12} \, .
$$
Since $\mathrm{L}$-operators are constructed from generators of
the symmetry algebra the previous relation implies
\begin{equation}
\label{intertwineR2} \mathrm{R}^{2}_{12} \cdot \left( \,
\mathrm{S}^+_1(\ell_1) + \mathrm{S}^+_2(\ell_2)
 \, \right) = \left( \, \mathrm{S}^{+}_1(\ell_1 - \alpha) +
\mathrm{S}^{+}_2(\ell_2 + \alpha) \, \right) \cdot
\mathrm{R}^{2}_{12}
\end{equation}
where we show explicitly the spin parameter of generators
$\mathrm{S}^{+}_k(\ell) = z_k^2\partial_k-2\ell z_k$ and introduce
the notation $\alpha = \frac{u_2 - v_2}{2}$. This relation and the
commutativity  of $\mathrm{R}^{2}_{12}$ with $z_{2}$ allow to
compute the action of $\mathrm{R}^{2}_{12}(u)$ on the generating
function $(1-x z_1)^{2\ell_1}$. The whole calculation is divided
into three steps: first we transform the initial expression using
commutativity $\mathrm{R}^{2}_{12}\, z_{2} = z_{2}\,
\mathrm{R}^{2}_{12}$
$$
\mathrm{R}^{2}_{12} \, (1-x z_1)^{2\ell_1} = (1-x z_{2})^{-2
\ell_2} \cdot \mathrm{R}^{2}_{12} \cdot (1-x z_1 )^{2\ell_1} (1-x
z_{2})^{2\ell_2}\ ,
$$
then use the representation for the generating function (\ref{gen}) and
use also the intertwining relation (\ref{intertwineR2})
$$
\mathrm{R}^{2}_{12}\cdot \mathrm{exp}\, \,x
\left(\mathrm{S}^{+}_{\ell_1}+\mathrm{S}^{+}_{\ell_2}\right) \cdot
1 = \mathrm{exp}\,\, x \left( \mathrm{S}^{+}_{\ell_1-\alpha} +
\mathrm{S}^{+}_{\ell_2+\alpha}\right) \cdot \mathrm{R}^{2}_{12}
\cdot 1\ ,
$$
calculate the emerging  constant $\mathrm{C} = \mathrm{R}^{2}_{12}
\cdot 1 = \frac{\Gamma(u_1+1)}{\Gamma(u_1-u_2+1)}$ and use
(\ref{gen}) once more
$$
\mathrm{exp}\,\, x \left( \mathrm{S}^{+}_{\ell_1-\alpha} +
\mathrm{S}^{+}_{\ell_2+\alpha}\right) \cdot 1 = (1-x z_1)^{2\ell_1
- 2\alpha}(1-x z_{2})^{2\ell_2 + 2\alpha}\ .
$$
Collecting everything  we arrive at
$$
\mathrm{R}^{2}_{12}\, (1-x z_1)^{2\ell_1} = \mathrm{C} \cdot (1-x
z_1)^{2\ell_1-2\alpha} \cdot (1- x z_{2})^{2\alpha} \, .
$$
Going back to (\ref{Qgenerat}) we fix $v_2=0\ ;\
\ell_1=\ell_2=\ell$, use the specification to arbitrary
sites $k, k+1$
\begin{equation} \label{R2polynom}
\mathrm{R}^{2}_{k k+1}(u)(1-x_k z_k)^{2\ell} =
\frac{\Gamma(-\ell+u)}{\Gamma(-2\ell)}\cdot (1-x_k z_k)^{\ell-u}
\cdot (1- x_k z_{k+1})^{\ell+u}
\end{equation}
and obtain the closed expression for the action of the considered operator
on the generating function
$$
\mathrm{R}^{2}_{1 2} (u) \, \mathrm{R}^{2}_{2 3} (u) \, \cdots \,
\mathrm{R}^{2}_{N 0} (u) \cdot (1-x_{1}z_{1})^{2\ell}
\cdots(1-x_{N} z_{N})^{2\ell} =
$$
$$
= \frac{\Gamma^N(-\ell+u)}{\Gamma^N(-2\ell)}\cdot(1-x_{1}
z_{1})^{\ell-u}(1-x_{1} z_{2})^{\ell+u}\cdots(1-x_{N}
z_{N})^{\ell-u}(1-x_{N} z_{0})^{\ell+u}.
$$
In order to obtain the wanted formula for the action of $\mathrm{Q}_2(u)$ on
the generating function it remains to put $z_0 = z_1$ and to perform the
cyclic shift $\mathrm{P}$.

The evident drawback of this formula is the presence of
$\Gamma$-functions. To improve this we introduce the
renormalized operator
\begin{equation} \label{Q}
\mathrm{Q}(u) = \frac{\Gamma^N(-2\ell)}{\Gamma^N(-\ell+u)}\cdot
\mathrm{Q}_2(u)\ .
\end{equation}
Its action on the generating function looks  simpler
\begin{equation}
\mathrm{Q}(u): (1-x_{1} z_{1})^{2\ell}\cdots(1-x_{N}
z_{N})^{2\ell} \mapsto \label{Q24}
\end{equation}
$$
\mapsto (1-x_{1} z_{N})^{\ell-u}(1-x_{1} z_{1})^{\ell+u}\cdot
\cdots(1-x_{N} z_{N-1})^{\ell-u}(1-x_{N} z_{N})^{\ell+u} \,
$$
The operator $\mathrm{Q}(u)$ has two further advantages. It is
normalized in a such way that $\mathrm{Q}(u): 1 \mapsto 1$ and it
is evident that it maps any monomial $z_1^{k_1}\cdots z_N^{k_N}$
to polynomial with respect to variables $z_1,\cdots, z_N$ and the
spectral parameter $u$ so that it maps polynomials in $z_1 \cdots
z_N$ into polynomials in $u,z_1 \cdots z_N$
$$
\mathrm{Q}(u) :\ \mathbb{C}[z_1\cdots z_N] \mapsto
\mathbb{C}[u,z_1 \cdots z_N]\ .
$$
This property guarantees that the operator $\mathrm{Q}(u)$ has
polynomial in $u$ eigenvalues and that the polynomials $Q_k(u)$
appearing in the algebraic Bethe ansatz approach~(\ref{Qk}) are
just the eigenvalues of the Q-operator.

Finally it is easy to check that we obtain the canonical form of
the Baxter equation for this improved operator
$$
\mathrm{t}(u)\mathrm{Q}(u) = (u-\ell)^N\cdot\mathrm{Q}(u+1)+
(u+\ell)^N\cdot\mathrm{Q}(u-1)\ .
$$
Note that this $\mathrm{Q}(u)$ coincides explicitly with the
$\mathrm{Q}$-operator constructed by another method in~\cite{SDQ}.

\section{Global objects: commuting transfer matrices}

\label{global} \setcounter{equation}{0}

In   section 2 we have introduced and investigated local
operators which concern only one site of the spin chain. Now we turn
to the description of the whole system. We are going to construct
various generating functions of commuting operators,  transfer matrices and
Baxter $\mathrm{Q}$-operators,  from general
$\R$-operators studied above.

\subsection{General transfer matrix and factorization into
$\mathrm{Q}$-operators}

\label{gen_tm}

It is of interest to generalize the previous construction of
the transfer matrix $\mathrm{t}(u)$.
The construction of the  general
transfer matrix  $\mathrm{T}(u)$ substitutes in formula~(\ref{tmat})
 $\mathrm{L}_k(u)$ as
local operators by $\R_{k0}(u|\ell,s)$ acting in the tensor product
of quantum space $\mathbb{U}_{-\ell}$ and auxiliary space
$\mathbb{U}_{-s}$. The  trace is taken over the generic infinite-dimensional
auxiliary space
\begin{equation} \label{T}
\mathrm{T}_{s}(u) = \tr_{0} \R_{10} (u|\ell, s)\,\R_{20} (u|\ell,
s) \cdots \R_{N0} (u|\ell, s)
\end{equation}
At fixed spin $\ell$ the  free parameters
in the general transfer matrix $\mathrm{T}_s(u)$ are the spectral
parameter $u$ and the spin parameter $s$ in the auxiliary space.
We recall the relation to our four-parameter notation (\ref{param}),
$$\R_{k0}(u-v|\ell,s)=  \R_{k0} (u_1,u_2;v_1,v_2), $$
$$ u_1 = u-\ell -1, u_2 = u+\ell, v_1 = v-s-1, v_2 = v+ s .$$
In this notation the above definition can be rewritten as
\begin{equation}
\mathrm{T}_{s}(u-v) = \tr_{0} \R_{10}(u_{1},u_2|v_1,v_{2})\cdots
\R_{N0}(u_{1},u_2|v_1,v_{2}).
\end{equation}
The general transfer matrix has the remarkable factorization
property
\begin{equation}\label{Factor1}
\mathrm{P}\cdot\mathrm{T}_{s} (u-v) =
\mathrm{Q}_2(u-v_2)\,\mathrm{Q}_1(u-v_1) =
\mathrm{Q}_1(u-v_1)\,\mathrm{Q}_2(u-v_2)
\end{equation}
where operators $\mathrm{Q}_1$ and $\mathrm{Q}_2$ are transfer
matrices constructed from operators $\R^{1}_{k0}$ and $\R^2_{k0}$
\begin{equation}\label{Q1}
\mathrm{Q}_1(u-v_1) = \tr_{0}\, \R^{1}_{10}(u_{1}|v_1,u_{2})\cdots
\R^{1}_{N0}(u_{1}|v_1,u_{2})\,,
\end{equation}
\begin{equation}\label{Q2}
\mathrm{Q}_2(u-v_2) = \tr_{0}\, \R^{2}_{10}(u_{1},u_2|v_{2})\cdots
\R^{2}_{N0}(u_{1},u_2|v_{2})\,,
\end{equation}
and the operator $\mathrm{P} =
\mathrm{P}_{12}\mathrm{P}_{13}\cdots\mathrm{P}_{1N}$ is the cyclic
permutation along the closed chain.

Note that the dependence on parameters $v_1$ and $v_2$ results in
a simple shift of spectral parameter,  $u\to u-v_1$ in the first
operator $\mathrm{Q}_1$ and $u\to u-v_2$ in the second one.
Eliminating the redundant shift of spectral parameter ($u-v \to
u$) we have
$$
\mathrm{P}\cdot\mathrm{T}_{s} (u) = \mathrm{Q}_2 (u - s) \,
\mathrm{Q}_1 (u + s + 1) = \mathrm{Q}_1 (u + s + 1) \,
\mathrm{Q}_2 (u - s)  \, .
$$
The factorization~(\ref{Factor1}) of transfer matrices generalizes
the corresponding properties of its building blocks
\begin{equation}
\R_{12}(u_{1},u_2|v_{1},v_2) = \mathrm{P}_{12}
\,\mathrm{R}^{1}_{12}(u_1|v_1,u_2)\,\mathrm{R}^{2}_{12}(u_1,u_2|v_{2})
\end{equation}
The global factorization follows from the local three term
relations~(\ref{R2RR}) and~(\ref{R1RR}). For doing the proof
we start form the
relation~(\ref{R2RR}) and rewrite it for the operators with
permutations included $\R_{ik} = \mathrm{P}_{ik}\,\mathrm{R}_{ik}$
$$
\R^2_{23}(v_1,v_2|w_2) \R_{13}(u_{1},u_2|w_1,w_2)
\R_{12}(u_{1},u_2|v_1,v_{2}) =
$$
$$
= \R_{12}(u_{1},u_2|w_1,v_{2}) \R_{13}(u_1,u_{2}|v_1,w_{2})
\R^2_{23}(v_{1},v_{2}|w_2).
$$
Now we choose the first space to be the local quantum space
$\mathbb{U}_{-\ell}$ in site $k$, the second space to be the auxiliary
space $\mathbb{U}_{-s}\sim \mathbb{C}[z_0]$ and the third space to be
a second copy of the auxiliary space $\mathbb{U}_{-s}\sim
\mathbb{C}[z_{0^{\prime}}]$
$$
\R^2_{00^{\prime}}(v_1,v_2|w_2)
\R_{k0^{\prime}}(u_{1},u_2|w_1,w_2)
\R_{k0}(u_{1},u_2|v_1,v_{2}) =
$$
\be \label{rf2'} =\R_{k0}(u_{1},u_2|w_1,v_{2})
\R_{k0^{\prime}}(u_1,u_{2}|v_1,w_{2})
\R^2_{00^{\prime}}(v_{1},v_{2}|w_2)\, . \ee
Recall that
$\R$-operators simplify when some of the their parameters coincide
$$
\R(u_1 , u_2| u_1 , v_2) = \R^{2}(u_1,u_2|v_{2}) \, , \qquad
\R(u_1 , u_2| v_1 , u_2) = \R^{1}(u_1|v_1,u_2)\,,
$$
so that specifying the parameters $w_1$ and $w_2$ as
$w_1=u_1$ and $w_2=u_2$ we obtain the local intertwining relation
with the operator $\R^2_{00^{\prime}}(v_{1},v_{2}|u_2)$
$$
\R^2_{00^{\prime}}(v_1,v_2|u_2)\cdot \mathrm{P}_{k0^{\prime}}
\cdot \R_{k0}(u_{1},u_2|v_1,v_{2}) =
\R^{2}_{k0}(u_{1},u_2|v_{2})\cdot
\R^1_{k0^{\prime}}(u_1|v_1,u_{2})\cdot
\R^2_{00^{\prime}}(v_{1},v_{2}|u_2)
$$
leading in the standard way to the relation for the transfer
matrices
\begin{equation}\label{factorI}
\tr_{0^{\prime}}
\left[\mathrm{P}_{10^{\prime}}\cdots\mathrm{P}_{N0^{\prime}}\right]\cdot
\tr_{0} \left[\R_{10}(u_{1},u_2|v_1,v_{2})\cdots
\R_{N0}(u_{1},u_2|v_1,v_{2})\right] =
\end{equation}
$$
= \tr_{0} \left[\R^{2}_{10}(u_{1},u_2|v_{2})\cdots
\R^{2}_{N0}(u_{1},u_2|v_{2})\right]\cdot \tr_{0^{\prime}} \left[
\R^1_{10^{\prime}}(u_1|v_1,u_{2})\cdots
\R^1_{N0^{\prime}}(u_1|v_1,u_{2})\right].
$$
We see that the general transfer matrix constructed from operators
$\R_{k0}(u_{1},u_2|v_1,v_{2})$ factorizes into the product of two
transfer matrices constructed from
$\R^{2}_{k0}(u_{1},u_2|v_{2})$ and
$\R^1_{k0^{\prime}}(u_1|v_1,u_{2})$. This factorization for global
objects follows in a clear and direct way from the local relations
for their building blocks. The similar proof of the second
factorization is given in Appendix A.

The commutativity properties of the different general transfer
matrices can be summarized as follows
$$
[ \, \mathrm{T}_{s}(u) , \mathrm{Q}_k (v) \, ] = 0 \ \ ;\ \ [ \,
\mathrm{Q}_i (u) , \mathrm{Q}_k (v) \, ] = 0 \ \ ;\ \ [ \,
\mathrm{P} , \mathrm{Q}_{k}(u) \, ] = 0 \ \ ;\ \ i,k = 1,2\,.
$$
The direct consequence of the two factorizations (\ref{factorI})
and (\ref{factorII}) is the commutativity of the transfer matrices
$\mathrm{Q}_1$ and $\mathrm{Q}_2$. However it is more instructive
to derive commutativity from local intertwining relations. In
Appendix A for completeness we list the necessary relations.

Summarizing  we have deduced factorization and commutativity
properties for the transfer matrices with infinite-dimensional
auxiliary spaces starting  only from local relations~(\ref{RRR}),
(\ref{R1RR}), (\ref{R2RR}).

The transfer matrices $\mathrm{Q}_1$ and $\mathrm{Q}_2$
constructed from operators $\R^1$ and $\R^2$ have all properties
of the $\mathrm{Q}$-operators, introduced by
R.Baxter~\cite{Baxter} and the second one coincides with the
operator $\mathrm{Q}_2(u)$ constructed previously "by hand". For
this reason we denote these transfer operators by $\mathrm{Q}_k$.

From the list of the defining properties for
$\mathrm{Q}$-operators given in Introduction all commutativity
properties are proven already so that we shall focus in the
following on the Baxter equation.

In section \ref{Qbyhand} we have already constructed
$\mathrm{Q}_2$ (\ref{Q_2}) "by hand" as particular solution of the
Baxter equation. In Appendix B it is shown that the trace over
infinite-dimensional space in (\ref{Q2}) can be calculated
explicitly due to specific form of $\mathrm{R}^2$ and it produces
exactly (\ref{Q_2}), i.e. the two operators coincide. Thus we have
an explicit useful formulae for the action of $\mathrm{Q}_2$ on
polynomials. Using the slightly modified argumentation of section
\ref{Qbyhand} it is easy to derive Baxter equation for
$\mathrm{Q}_2$ in the form~(\ref{Q2}) directly from the defining
relation of the operator $\mathrm{R}^2$ and the same is true for
$\mathrm{Q}_1$ in the form~(\ref{Q1}).

For completeness we present another
derivation~\cite{DeMa,Shortcut} of the Baxter equation which is
deeper and shows the origin of this equation: at integer values of
$2s$ there appears a finite-dimensional invariant subspace inside
the infinite-dimensional representation of the algebra $s\ell_2$ in
auxiliary space and this finally  results in the Baxter equation.
The general transfer matrix is constructed from general
$\R$-operators so that  we have in principle to consider at first
the restriction to the finite-dimensional invariant subspaces of
these local building blocks. All this will be discussed in detail
in the next section. Here we simply use the needed formulae
postponing their proof to the more appropriate moment.

\subsection{Baxter equation and determinant formula}
\label{det-formula}

In the previous section we have introduced the general transfer matrix
(\ref{T}) for generic spin parameter $s$ in the auxiliary space
$\mathbb{U}_{-s}$. Now we are going to chose this parameter to be
$s = \frac{n}{2}, \, n =0, 1 , 2 , \ldots$ As we have mentioned in
\ref{L-operators} in this case the auxiliary module
$\mathbb{U}_{-\frac{n}{2}}$ is now reducible. It is useful to the
introduce operator $\mathcal{D} = \dd^{n+1}$ intertwining
generators of the algebra with parameters $\frac{n}{2}$ and
$-\frac{n}{2}-1$
\begin{equation} \label{intertwine} \mathcal{D}
\cdot \mathrm{S}_{\pm}\left({\textstyle\frac{n}{2}}\right) =
\mathrm{S}_{\pm}\left(-1-{\textstyle\frac{n}{2}}\right) \cdot
\mathcal{D} \ \ ;\ \ \mathcal{D} \cdot
\mathrm{S}\left({\textstyle\frac{n}{2}}\right) =
\mathrm{S}\left(-1-{\textstyle\frac{n}{2}}\right) \cdot
\mathcal{D}\ .
\end{equation}
It is easy to see that image and kernel of the operator
$\mathcal{D}$ are invariant subspaces. The kernel is the
$(n+1)$-dimensional space $\mathbb{V}_n$ with the basis $\{1,
z,\cdots z^{n}\}$ and the image is the infinite-dimensional space with
basis $\{1, z, z^{2},  \ldots \}$ which is also irreducible.
The operator $\mathcal{D}$ maps the reducible module with lowest
weight $-\frac{n}{2}$ into the irreducible module with lowest
weight $\frac{n}{2}+1$\,: $\mathrm{Im}\mathcal{D} =
\mathbb{U}_{\frac{n}{2} + 1}\ ,\ \mathrm{Ker}\mathcal{D} =
\mathbb{V}_n$
$$
\mathbb{U}_{-\frac{n}{2}} \xrightarrow{\mathcal{D}}
\mathbb{U}_{\frac{n}{2} + 1} \ \ ;\ \ \mathbb{V}_n
\xrightarrow{\mathcal{D}} 0\,,
$$
where the irreducible module $\mathbb{U}_{\frac{n}{2}+1}$ is
isomorphic to the quotient module $\mathbb{U}_{\frac{n}{2}+1}
\approx \mathbb{U}_{-\frac{n}{2}}/\mathbb{V}_n$ and the
isomorphism is induced by intertwining operator $\mathcal{D}$. As
the consequence the trace over $\mathbb{U}_{-\frac{n}{2}}$ splits
into traces over finite-dimensional $\mathbb{V}_n$ and
infinite-dimensional $\mathbb{U}_{\frac{n}{2}+1}$.

Applying this statement to the general transfer matrix we obtain
its decomposition into the transfer matrix with finite-dimensional
auxiliary space and the general transfer matrix with the other
spin parameter:
\begin{equation} \label{transfer}
\mathrm{T}_{\frac{n}{2}}(u) = \mathrm{t}_{n}(u)+
\mathrm{T}_{-\frac{n}{2} - 1}(u) \, .
\end{equation}
Here in $\mathrm{t}_{n}(u)$ the trace is taken over $\mathbb{V}_n$
and it represents the generalization of the ordinary transfer matrix
$\mathrm{t}(u)$ considered in the section \ref{tm}. \be
\label{t_0} \mathrm{t}_{n}(u) = \tr \, \mathbf{R}_{10}\left(u|\ell
, {\textstyle\frac{n}{2}} \, \right) \,
\mathbf{R}_{20}\left(u|\ell , {\textstyle\frac{n}{2}}\, \right) \,
\ldots \, \mathbf{R}_{N0}\left(u|\ell , {\textstyle\frac{n}{2}} \,
\right) \, . \ee $\mathbf{R}_{k0}(u|\ell , \frac{n}{2})$ is the
restriction of the general $\R$-operator to the space
$\mathbb{U}_{-\ell} \otimes \mathbb{V}_n$. In  section
\ref{R->L} we shall calculate explicitly such restrictions for
one- and two-dimensional auxiliary spaces. Using formula
(\ref{R_0}) we have \footnote{
In this subsection we change normalization of
$\mathrm{R}$-operators
$\mathrm{R}(u_1 , u_2|v_1 , v_2) \to (-1)^{u_1-v_1}\mathrm{R}(u_1 , u_2|v_1 , v_2)$
and $\mathrm{R}^1(u_1|v_1 , v_2) \to (-1)^{u_1-v_1}\mathrm{R}^1(u_1|v_1 , v_2)$
in order to obtain Baxter relation in the standard form.
In the other parts of the paper we do not retain this normalization factor since
it can be restored easily. } 

\begin{equation}\label{t_1}
\mathrm{t}_{0} (u) = (-1)^{N(u-\ell)}
\frac{\Gamma^N\left(-\ell+u\right)}{\Gamma^N\left(-\ell-u\right)}
\cdot \II \, ,
\end{equation} 
and (\ref{R_1/2}) allows to connect $\mathrm{t}_{1} (u)$ with the
standard transfer matrix $\mathrm{t}(u)$ considered in the section
\ref{tm}

$$
\mathrm{t}_{1} \left( u - {\textstyle\frac{1}{2}} \right) = (-1)^{N(u-\ell)}
\frac{\Gamma^N\left(-\ell+u\right)}{\Gamma^N\left(-\ell-u\right)}\cdot
\frac{\mathrm{t} (u)}{(u_1 u_2)^N} \, .
$$

The trace in $\mathrm{T}_{-\frac{n}{2} - 1}(u) $ is taken over
$\mathbb{U}_{\frac{n}{2}+1}$. The operator $\mathcal{D}$ is
invertible on this space and using the invariance of the trace
with respect to similarity transformation we can write
$$
\mathrm{T}_{-\frac{n}{2} - 1 }(u) = \tr \mathcal{D} \,
\R_{10}\left(u|\ell , {\textstyle\frac{n}{2}} \,\right) \,
\R_{20}\left(u|\ell , {\textstyle\frac{n}{2}} \,\right) \, \ldots
\, \R_{N0}\left(u|\ell , {\textstyle\frac{n}{2}} \,\right) \,
\mathcal{D}^{-1} \, .
$$
Using the similarity transformation in the $\mathrm{RLL}$-relation
(\ref{RLL}) and the intertwining property (\ref{intertwine}) of
the operator $\mathcal{D}$ one obtains
$$
\mathrm{T}_{-\frac{n}{2} - 1 }(u) = \tr \, \R_{10}\left(u|\ell , -
{\textstyle\frac{n}{2}} - 1 \,\right) \, \R_{20}\left(u|\ell , -
{\textstyle\frac{n}{2}} - 1 \,\right) \, \ldots \,
\R_{N0}\left(u|\ell , - {\textstyle\frac{n}{2}} - 1 \,\right) \,
$$
that is in agreement with definition of (\ref{T}).

The factorization properties of the general transfer matrix
(\ref{Factor1}) lead then from (\ref{transfer}) to the determinant
formula
$$
\mathrm{P}\cdot\mathrm{t}_{n}(u) = \mathrm{Q}_1\left(u +
{\textstyle\frac{n}{2}} + 1\right)\,\mathrm{Q}_2\left(u -
{\textstyle\frac{n}{2}}\right) - \mathrm{Q}_1 \left(u -
{\textstyle\frac{n}{2}}\right) \, \mathrm{Q}_2 \left(u +
{\textstyle\frac{n}{2}} + 1\right) =
$$
$$
= \left|\begin{array}{cc}
\mathrm{Q}_1\left(u+ \frac{n}{2} +1\right) & \mathrm{Q}_2\left(u+ \frac{n}{2} +1\right) \\
\mathrm{Q}_1\left(u - \frac{n}{2}\right)& \mathrm{Q}_2\left(u -
\frac{n}{2}\right)\end{array}\right|
$$
allowing to express the transfer matrix with finite-dimensional
auxiliary space in terms of Baxter $\mathrm{Q}$-operators. Then we
proceed to establish a set of relations which are linear in
transfer matrices and $\mathrm{Q}$-operators~\cite{DeMa}. Let us
consider the following determinant which is zero due to
equality of  two columns
$$
\left|\begin{array}{ccc}
\mathrm{Q}_1(a) & \mathrm{Q}_2(a) & \mathrm{Q}_k(a)\\
\mathrm{Q}_1(b) & \mathrm{Q}_2(b) & \mathrm{Q}_k(b)\\
\mathrm{Q}_1(c) & \mathrm{Q}_2(c) &\mathrm{Q}_k(c)
\end{array}\right| = 0 \ \ ;\ \ k = 1 , 2\,.
$$
The decomposition with respect to the third column results in
$$
\left|\begin{array}{cc}
\mathrm{Q}_1(b) & \mathrm{Q}_2(b) \\
\mathrm{Q}_1(c)& \mathrm{Q}_2(c)\end{array}\right|\cdot
\mathrm{Q}_k(a) - \left|\begin{array}{cc}
\mathrm{Q}_1(a) & \mathrm{Q}_2(a) \\
\mathrm{Q}_1(c)& \mathrm{Q}_2(c)\end{array}\right|\cdot
\mathrm{Q}_k(b) +\left|\begin{array}{cc}
\mathrm{Q}_1(a) & \mathrm{Q}_2(a) \\
\mathrm{Q}_1(b)& \mathrm{Q}_2(b)\end{array}\right|\cdot
\mathrm{Q}_k(c) = 0 \, .
$$
Specifying parameters
$$
a = u + {\textstyle\frac{n}{2}} + 1 \ ; \ b = u -
{\textstyle\frac{n}{2}}\ ; \ c = u -m-{\textstyle\frac{n}{2}} - 1
$$
we see that the quadratic determinants turns into
transfer matrices and we get the set of relations at $n,m = 0 ,1 , 2
, \ldots$
$$ \mathrm{t}_{m}\left(u-1-{\textstyle\frac{n+m}{2}}\right)\cdot
\mathrm{Q}_k\left(u+1+{\textstyle\frac{n}{2}}\right)
-\mathrm{t}_{n+m+1}\left(u-{\textstyle\frac{m+1}{2}}\right)\cdot
\mathrm{Q}_k\left(u-{\textstyle\frac{n}{2}}\right) +
$$
$$
+ \mathrm{t}_{n}\left(u \right)\cdot
\mathrm{Q}_k\left(u-1-m-{\textstyle\frac{n}{2}}\right) = 0
$$
Let us assign in the previous relation $n = m = 0$
\begin{equation}\label{B}
\mathrm{t}_{1}\left(u-{\textstyle\frac{1}{2}}\right)\cdot
\mathrm{Q}_k\left(u\right) = \mathrm{t}_0\left(u-1\right)\cdot
\mathrm{Q}_k\left(u+1\right)+ \mathrm{t}_{0}\left(u\right)\cdot
\mathrm{Q}_k\left(u-1\right) \, .
\end{equation}
Taking into account the expressions for transfer matrices with 1-dimensional
and 2-dimensional auxiliary spaces (\ref{t_0}), (\ref{t_1}) we obtain
Baxter relations
$$
\mathrm{t}\left(u\right)\cdot \mathrm{Q}_k\left(u\right) =
\mathrm{Q}_k\left(u+1\right) + (u_1 u_2)^N\cdot
\mathrm{Q}_k\left(u-1\right) \, .
$$

\subsection{Explicit formulae for action on polynomials}
\label{explicit}

Above we have given the mostly algebraic construction of
$\mathrm{Q}$-operators acting in infinite-dimensional quantum
space in the case of generic spin $\ell\in \mathbb{C}$. Now we are
going to consider explicit formulae for the action of these
operators on polynomials.

We start from the uniform expressions for the operators
$\mathrm{Q}_{1}$ and $\mathrm{Q}_{2}$
$$
\mathrm{Q}_{1}(u) =
\tr_{\mathbb{V}_{0}}\mathrm{P}_{10}\,\mathrm{R}_1\left(z_{01}\dd_0\right)\cdots
\mathrm{P}_{N0}\,\mathrm{R}_1\left(z_{0N}\dd_0\right)\,,
$$
$$
\mathrm{Q}_{2}(u) =
\tr_{\mathbb{V}_{0}}\mathrm{P}_{10}\,\mathrm{R}_2\left(z_{10}\dd_1\right)\cdots
\mathrm{P}_{N0}\,\mathrm{R}_2\left(z_{N0}\dd_N\right)\,,
$$
where we use the following functions of its operator arguments
\begin{equation}\label{Rfunc}
\mathrm{R}_1(x) = \frac{\Gamma\left(x-2\ell\right)}
{\Gamma\left(x+1-\ell-u\right)}\ \ \ ;\ \ \ \mathrm{R}_2(x) =
\frac{\Gamma\left(x+u-\ell\right)} {\Gamma\left(x-2\ell\right)}\,.
\end{equation}
This expression suggests the  derivation of the explicit
expression for the action on polynomials.
We use
a simple trick for separating the  dependence on spectral
parameter $u$ and spin $\ell$. The
 evident formula
$$
\left.\Phi(\lambda\partial_{\lambda})\right|_{\lambda=1}\cdot\lambda^{x}
= \Phi(x)\,
$$
allows to extract the dependence on $u$ and $\ell$  into
operators involving $\lambda\partial_{\lambda}$
$$
\mathrm{Q}_{1}(u) =
\left.\mathrm{R}_1(\lambda_1\partial_{\lambda_1})\cdots
\mathrm{R}_1(\lambda_N\partial_{\lambda_N})\right|_{\lambda=1}\cdot
\tr_{\mathbb{V}_{0}}\mathrm{P}_{10}\lambda_1^{z_{01}\dd_0}\cdots
\mathrm{P}_{N0}\lambda_N^{z_{0N}\dd_0}\,,
$$
$$
\mathrm{Q}_{2}(u) =
\left.\mathrm{R}_2(\lambda_1\partial_{\lambda_1})\cdots
\mathrm{R}_2(\lambda_N\partial_{\lambda_N})\right|_{\lambda=1}\cdot
\tr_{\mathbb{V}_{0}}\mathrm{P}_{10}\lambda_1^{z_{10}\dd_1}\cdots
\mathrm{P}_{N0}\lambda_N^{z_{N0}\dd_N}\,.
$$
After this transformation we can focus on the problem of
calculation of the traces in these two basic expressions.
For the
 calculation of the trace in the space of polynomials
$\mathbb{C}[z_0]$ we adopt the standard procedure. We use the monomial
basis $\mathbf{e}_k = z_0^k$ in the space $\mathbb{C}[z_0]$ and
the standard definition for matrix of an operator $ \mathrm{A}
\mathbf{e}_i = \sum_{k} \mathbf{e}_k \mathrm{A}_{ki}$ and then
calculate the trace of operator $\mathrm{A}$ as the sum of
diagonal matrix elements $\tr \mathrm{A} = \sum_{k}
\mathrm{A}_{kk}$.

Let us start with the second operator because this trace is the
simplest one. It is the special case of the general situation
considered in Appendix B and the result of calculation of the
trace is
\begin{equation}\label{tr21}
\tr_{\mathbb{V}_{0}}\mathrm{P}_{10}\lambda_1^{z_{10}\dd_1}\cdots
\mathrm{P}_{N0}\lambda_N^{z_{N0}\dd_N} = \left.\mathrm{P}\cdot
\lambda_1^{z_{12}\dd_1}\,\lambda_2^{z_{23}\dd_2}\cdots
\lambda_N^{z_{N0}\dd_N}\right|_{z_0=z_1}
\end{equation}
The result of the action of the obtained operator on a function
$\Psi(z_1 , \ldots , z_N) = \Psi(\vec{z})$ can be expressed in a
compact matrix form
$$
\left.\lambda_1^{z_{12}\dd_1}\,\lambda_2^{z_{23}\dd_2}\cdots
\lambda_N^{z_{N0}\dd_N}\right|_{z_0=z_1}\cdot\Psi(\vec{z}) =
\Psi(\Lambda \vec{z})\,,
$$
where
$$
\Lambda = \begin{pmatrix}
\lambda_1 & 1-\lambda_1 & 0 & 0 & \hdotsfor{2} & 0 \\
0 & \lambda_2 & 1-\lambda_2 & 0 & \hdotsfor{2} & 0 \\
0 & 0 & \lambda_3 & 1-\lambda_3 & \hdotsfor{2} & 0 \\
\hdotsfor{7}  \\
0 & 0 & \hdotsfor{3} & \lambda_{N-1} & 1-\lambda_{N-1} \\
1-\lambda_{N} & 0 & 0 & 0 & \cdots & 0 & \lambda_{N}
\end{pmatrix} \ \ ;\ \ \vec{z} = \begin{pmatrix}
z_1\\ z_2\\ z_3\\ \hdotsfor{1} \\ \hdotsfor{1} \\ z_{N}
\end{pmatrix}
$$
Now we turn to the first operator $\mathrm{Q}_1$. The
calculation of the trace is based on the formula
\begin{equation} \label{sSum}
\sum_{k=0}^{\infty}\frac{1}{k!} \dd^k_{0} \cdot\left(a+ b\cdot
z_0\right)^k \Phi(z_0)\biggl|_{z_0=0} = \frac{1}{1-b}\cdot
\Phi\left(\frac{a}{1-b}\right)\,,
\end{equation}
which is proven in Appendix B.
Also the calculation of the involved trace
$
\tr_{\mathbb{V}_{0}}\mathrm{P}_{10}\lambda_1^{z_{01}\dd_0}\cdots
\mathrm{P}_{N0}\lambda_N^{z_{0N}\dd_0}
$
is given in Appendix B with the result
\be \label{TraceExpl}
\tr_{\mathbb{V}_{0}}\mathrm{P}_{10}\lambda_1^{z_{01}\dd_0}\cdots
\mathrm{P}_{N0}\lambda_N^{z_{0N}\dd_0}\cdot \Psi(\vec{z}) =
\frac{1}{ 1 - \bar{\lambda}_1 \cdots \bar{\lambda}_N } \cdot
\Psi(\Lambda^{\prime -1}\, \vec{z} \, ) \,,
\ee
 where
$$
\Lambda^{\prime} = \begin{pmatrix}
1-\frac{1}{\lambda_1} & \frac{1}{\lambda_1} & 0 & 0 & \hdotsfor{2} & 0 \\
0 & 1-\frac{1}{\lambda_2} & \frac{1}{\lambda_2} & 0 & \hdotsfor{2} & 0 \\
0 & 0 & 1-\frac{1}{\lambda_3} & \frac{1}{\lambda_3} & \hdotsfor{2} & 0 \\
\hdotsfor{7}  \\
0 & 0 & \hdotsfor{3} & 1-\frac{1}{\lambda_{N-1}} & \frac{1}{\lambda_{N-1}} \\
\frac{1}{\lambda_{N}} & 0 & 0 & 0 & \cdots & 0 &
1-\frac{1}{\lambda_{N}}
\end{pmatrix} \,.
$$
As the result
we obtain the very similar formulae for the action of
the operators $\mathrm{Q}_k$ on polynomials
\begin{equation}\label{Q1generic}
\mathrm{Q}_{1}(u)\,\Psi(\vec{z}) =
\left.\mathrm{R}_1(\lambda_1\partial_{\lambda_1})\cdots
\mathrm{R}_1(\lambda_N\partial_{\lambda_N})\right|_{\lambda=1}\cdot
\frac{1}{ 1 - \bar{\lambda}_1 \cdots \bar{\lambda}_N } \cdot
\Psi(\Lambda^{\prime -1}\, \vec{z} \, ) \,,
\end{equation}
\begin{equation}\label{Q2generic}
\mathrm{Q}_{2}(u)\,\Psi(\vec{z}) =
\mathrm{P}\cdot\left.\mathrm{R}_2(\lambda_1\partial_{\lambda_1})\cdots
\mathrm{R}_2(\lambda_N\partial_{\lambda_N})\right|_{\lambda=1}\cdot
\Psi(\Lambda\vec{z})\,.
\end{equation}
The main difference is in the prefactor and the inversion of the matrix in the
first formula. To avoid misunderstanding we quote how the operator
$\mathrm{P}$ acts on the function:
$\mathrm{P}\Psi(z_1,z_2,\ldots,z_N) =
\Psi(z_N,z_1,\ldots,z_{N-1})$.

The formulae~(\ref{Q1generic}) and~(\ref{Q2generic}) are the starting
points for the derivation of various representations for the
$\mathrm{Q}$-operators. One kind of representations is in the form
of integral operator, which is obtained by using the simple
substitutions
\begin{equation}\label{rule1}
\left.\mathrm{R}_1(\lambda_k\dd_{\lambda_k})\right|_{\lambda_k =
1} \to \frac{1} {\Gamma(1+\ell-u)}\cdot
\int^{1}_{0}\mathrm{d}\lambda_k (1-\lambda_k)^{\ell-u}
\lambda_k^{-2\ell-1}\,,
\end{equation}
\begin{equation}\label{rule2}
\left.\mathrm{R}_2(\lambda_k\dd_{\lambda_k})\right|_{\lambda_k =
1} \to \frac{1}{\Gamma(-\ell-u)}\cdot
\int^{1}_{0}\mathrm{d}\lambda_k (1-\lambda_k)^{-\ell-u-1}
\lambda_k^{-\ell+u-1}
\end{equation}
so that we obtain the multiple integral representation for the
operator $\mathrm{Q}_1$
\begin{equation}
\left[\mathrm{Q}_{1}(u)\Psi\right](\vec{z}) = \frac{1}
{\Gamma^N(1+\ell-u)}\cdot \int^{1}_{0}\mathrm{d}\lambda_1
(1-\lambda_1)^{\ell-u} \lambda_1^{-2\ell-1}\cdots \label{Q-1}
\end{equation}
$$
\cdots \int^{1}_{0}\mathrm{d}\lambda_N
(1-\lambda_N)^{\ell-u}\lambda_N^{-2\ell-1} \, \frac{1}{ 1 -
\bar{\lambda}_1 \cdots \bar{\lambda}_N } \cdot \Psi(\Lambda^{-1}\,
\vec{z} \, ) \,,
$$
and for the operator $\mathrm{Q}_2$ this leads to the multiple
integral representation
\begin{equation}
\left[\mathrm{Q}_{2}(u)\Psi\right](z_1,\ldots, z_N) =
\frac{1}{\Gamma^N(-\ell-u)}\cdot \int^{1}_{0}\mathrm{d}\lambda_1
(1-\lambda_1)^{-\ell-u-1} \lambda_1^{-\ell+u-1}\cdots \label{Q-2}
\end{equation}
$$
\cdots \int^{1}_{0}\mathrm{d}\lambda_N
(1-\lambda_N)^{-\ell-u-1}\lambda_N^{-\ell+u-1} \,
\Psi\left(\lambda_1 z_{N}+\bar{\lambda}_1 z_1 , \lambda_2 z_{1}+
\bar{\lambda}_2 z_2 , \ldots , \lambda_N z_{N-1}+\bar{\lambda}_N
z_N \right).
$$
In fact the derivation of the rules~(\ref{rule1})
and~(\ref{rule2}) is reduced to the use of the beta-integral
representation for corresponding operator: in the case
$\mathrm{R}_2(\lambda\dd_{\lambda})$ it is the following chain of
transformations
$$
\mathrm{R}_2(\lambda\dd_{\lambda})\left.\Phi(\lambda)\right|_{\lambda
= 1} = \frac{1}{\Gamma(-\ell-u)}\cdot \int^{1}_{0}\mathrm{d}\alpha
(1-\alpha)^{-\ell-u-1}
\alpha^{-\ell+u-1}\cdot\alpha^{\lambda\dd_{\lambda}}
\left.\Phi(\lambda)\right|_{\lambda = 1}=
$$
$$
=\frac{1}{\Gamma(-\ell-u)}\cdot \int^{1}_{0}\mathrm{d}\alpha
(1-\alpha)^{-\ell-u-1} \alpha^{-\ell+u-1}\Phi(\alpha) \,,
$$
and on the last step for simplicity in~(\ref{rule2}) we change
$\alpha \to \lambda$ again.

We postpone to the next section the derivation of the useful
representation for the operator $\mathrm{Q}_1$ in the case of
half-integer spin. Here we add some comments about the operator
$\mathrm{Q}_2$ concerning the connection between the present
formula~(\ref{Q-2}) and representation~(\ref{Q24}) from
section~\ref{Qbyhand}. For the generating function
$(1-x_{1}z_{1})^{2\ell}\cdots(1-x_{N} z_{N})^{2\ell}$ in left hand
side of formula~(\ref{Q24}) the multiple integral (\ref{Q-2}) is
factorized to the product of simple integrals calculated
explicitly by using Feynman formula \be \label{F} \int_{0}^{1}
\mathrm{d}\alpha \, \alpha^{a-1}\,(1-\alpha)^{b-1}
\frac{1}{\bigl[\alpha
\mathrm{A}+(1-\alpha)\mathrm{B}\bigr]^{a+b}}\ =\
\frac{\Gamma(a)\Gamma(b)}{\Gamma(a+b)}\cdot\frac{1}{\mathrm{A}^a
\mathrm{B}^b}\, \ee so that finally we arrive up to normalization
at the right hand side of~(\ref{Q24}).


\section{Finite-dimensional representations}
\label{finiteDim}\setcounter{equation}{0}

In the previous section we have assumed the spin parameter $\ell$
to be a generic complex number and consequently the quantum space
of the model was infinite-dimensional. Under this assumptions we
have concentrated on the algebraic properties of
$s\ell_2$-invariant transfer matrices constructed from the
operators $\R^1$ and $\R^2$ and have demonstrated the key
properties that allow to call them $\mathrm{Q}$-operators:
commutativity and Baxter equation. Now we are going to consider
the special situation of a half-integer spin $\ell$. In this case
the infinite-dimensional representation becomes reducible and
there an invariant subspace appears which is the standard
finite-dimensional irreducible representation labeled by
half-integer spin $\ell$.

What happens if we put in the formulae obtained so far $\ell$ to
be half-integer and restrict all our operators to the
finite-dimensional quantum subspace? Let us turn to the
formula~(\ref{Q24}) since it is very useful for the discussion of
the finite-dimensional representations. We consider the simplest
example of  spin $\ell=\frac{1}{2}$ and two sites ($N=2$) for
illustration.
\begin{equation}\label{example}
\mathrm{Q}(u): (1-x_{1} z_{1})\cdot(1-x_{2} z_{2}) \mapsto
(1-x_{1} z_{2})^{\frac{1}{2}-u}(1-x_{1}
z_{1})^{\frac{1}{2}+u}\cdot (1-x_{2}z_{1})^{\frac{1}{2}-u}(1-x_{2}
z_{2})^{\frac{1}{2}+u} \, .
\end{equation}
The tensor product basis from two quantum spaces with bases
$\{1\, , z_1\}$ and $\{1\, , z_2\}$ is $\{1\, , z_1\, , z_2\, ,
z_1 z_2\}$.  From  formula~(\ref{example}) one easily obtains
the action of the operator $\mathrm{Q}(u)$ on these basis vectors
$$
z_1 \mapsto \left(\frac{1}{2}+u\right) z_1 +
\left(\frac{1}{2}-u\right) z_2 \ \ ;\ \ z_2 \mapsto
\left(\frac{1}{2}+u\right) z_2 + \left(\frac{1}{2}-u\right) z_1
$$
$$
1 \mapsto 1 \ \ ;\ \ z_1 z_2 \mapsto \left(\frac{1}{2}+2u^2\right)
z_1 z_2 + \left(\frac{1}{4}-u^2\right)\left(z_1^2+z_2^2\right)\ .
$$
Due to the presence of the term $\sim\left(z_1^2+z_2^2\right)$ the
operator $\mathrm{Q}(u)$ maps beyond the initial four-dimensional
space. It is clear that the same phenomenon occurs for any compact
spin $\ell = \frac{n}{2}\ ;\ n+1 \in \mathbb{N}$.

We see that the operator $\mathrm{Q}_2$ maps beyond the quantum space
and therefore we have to reconsider our construction in the case of
finite-dimensional representations. Our strategy will be the
following:
\begin{itemize}
\item From the very beginning we shall work with the restriction of
the general $\R$-operator to the invariant subspace appearing for
half-integer value of the spin.

\item We use this restricted $\R$-operator as building block for
the construction of the corresponding transfer matrix. At this
stage we must introduce a regularization because of divergence of
the trace over the infinite-dimensional space. In the case of
arbitrary spin $\ell\in \mathbb{C}$ the spin parameter $\ell$
itself plays the role of regulator but now we fix it to be
half-integer so that one needs a new regularization.

\item Using local relations we prove factorization of the
general transfer matrix into the product of the corresponding
$\mathrm{Q}$-operators.

\end{itemize}

The main difference to  the case of generic $\ell\in \mathbb{C}$
is that from  very beginning all operators are restricted to
invariant finite-dimensional subspace. Contrary to $\R^1$ and
$\R^2$, they map the finite-dimensional quantum space into itself. To
start with we consider in details two examples of such restriction
in the next subsection.

\subsection{Examples of restriction: one- and two-dimensional representations}

\label{R->L}

Now we are going to fill the gap left in the previous section
 and
consider two particular examples of restriction needed for the
derivation of the Baxter equation in the previous section
\ref{det-formula}. The $\R$-operator acts in the space
$\mathbb{U}_{-\ell_1}\otimes\mathbb{U}_{-\ell_2}$ and according to
(\ref{Rfact}) has the form
\begin{equation} \label{PR}
\R (u|\ell_1,\ell_2) =
\mathrm{P}_{12}\cdot\frac{\Gamma(z_{12}\dd_1-2\ell_2)}
{\Gamma(z_{12}\dd_2-\ell_1-\ell_2-u)}
\cdot\frac{\Gamma(z_{21}\dd_2-\ell_1-\ell_2+u)}
{\Gamma(z_{21}\dd_2-2\ell_2)}.
\end{equation}
Now we chose $\ell_2 = \frac{n}{2}$ and restrict the $\R$-operator
to the space $\mathbb{U}_{-\ell_1}\otimes\mathbb{V}_n$.

In this section we perform explicit and detailed calculations for
$n=0,1$. In the case $\ell_2 = 0$ we have
$$
\mathbb{U}_{-\ell_1}\otimes\mathbb{V}_0 \sim
\mathbb{U}_{-\ell_1}\otimes\C
$$
so that the restricted $\R$-operator acts on functions of the form
$\Psi(z_1,z_2) = \phi(z_1)$. The action of the first operator in
({\ref{PR}}) is simple due to independence on the variable $z_2$
$$
\mathrm{e}^{-z_1\dd_2}\frac{\Gamma(z_{2}\dd_2-\ell_1-\ell_2+u)}
{\Gamma(z_{2}\dd_2-2\ell_2)}\mathrm{e}^{z_1\dd_2} \cdot \phi(z_1)
= \phi(z_1)\cdot \frac{\Gamma(-\ell_1-\ell_2+u)} {\Gamma(-2\ell_2)}.
$$
Note that in the point $\ell_2 = 0$ the above expression turns
into zero. Therefore we have to introduce the regularization $2\ell_2
= -\varepsilon$. Then this expression has a simple  zero at
$\varepsilon \to 0$
$$
\phi(z_1)\cdot \frac{\Gamma\left(-\ell_1+u\right)}
{\Gamma(\varepsilon)} + O(\varepsilon^2).
$$
The introduction of regularization implies that we do not
substitute directly in (\ref{PR}) the value of compact spin but do
perform carefully the limiting procedure.
The action of the second operator is not trivial
$$
\mathrm{e}^{-z_2\dd_1}\frac{\Gamma(z_{1}\dd_1-2\ell_2)}
{\Gamma\left(z_{1}\dd_1-\ell_1-\ell_2-u\right)}
\mathrm{e}^{z_2\dd_1} \cdot \phi(z_1) = \sum_{k=0}^{\infty}
\frac{\phi^{(k)}(z_2)}{k!} \frac{\Gamma(k+\varepsilon)}
{\Gamma\left(k-\ell_1+\frac{\varepsilon}{2}-u\right)} \cdot
z_{12}^k
$$
but we  have actually to extract the singular part $\sim
\varepsilon^{-1}$ only.
In other words previous expression with needed
accuracy is equal to
$$
\phi(z_1)\cdot\frac{\Gamma(\varepsilon)}
{\Gamma\left(-\ell_1-u\right)} + O(\varepsilon^0).
$$
Now we see that the simple pole cancels out the simple zero
and we have for $\ell_1 = \ell$ and $\ell_2 \to 0$
$$
\R \left(u|\ell, 0 \right) \phi(z_1) =
\frac{\Gamma\left(-\ell+u\right)}
{\Gamma\left(-\ell-u\right)}\cdot \phi(z_1)
$$
or in the operator form
\begin{equation} \label{R_0}
\R\left(u|\ell, 0 \right)\left|_{\mathbb{V}_0}\right. =
\frac{\Gamma\left(-\ell+u\right)}
{\Gamma\left(-\ell-u\right)}\cdot \II\,.
\end{equation}
We would like to mention that the presence of divergences in the
considered operators is not a plague but rather an advantage allowing to
simplify the calculation considerably.

In the case $\ell_2 = \frac{1}{2}$ we have
$$
\mathbb{U}_{-\ell_1}\otimes\mathbb{V}_1 \sim
\mathbb{U}_{-\ell_1}\otimes\C^2
$$
and the $\mathrm{R}$-operator acts on the functions of the form
$$
\Psi(z_1,z_2) = \phi(z_1) + z_2 \psi(z_1).
$$
The action of the first operator in (\ref{PR}) is simple due to
special $z_2$-dependence
$$
\left[ \phi(z_1)+z_1\psi(z_1) + z_{12}\cdot\psi(z_1)\cdot
\left(-\ell_1-\textstyle{\frac{1}{2}}+u\right) \right]\cdot
\frac{\Gamma\left(-\ell_1-\textstyle{\frac{1}{2}}+u\right)}
{\Gamma(-1+\varepsilon)} + O(\varepsilon^2)
$$
As in the previous calculation we have to introduce regularization
$2\ell_2 = 1-\varepsilon$ in order to work with divergences. Again
due to $\Gamma(-1+\varepsilon)$ the previous expression has simple
zero at $\varepsilon\to 0$. The second operator in (\ref{PR})
commutes with $z_2$ so that we have to calculate its action on the
function which depends on variable $z_1$ only. The action of this
operator is not trivial
$$
\mathrm{e}^{-z_2\dd_1}\frac{\Gamma(z_{1}\dd_1-1+\varepsilon)}
{\Gamma\left(z_{1}\dd_1-\ell_1-\frac{1-\varepsilon}{2}-u\right)}
\mathrm{e}^{z_2\dd_1} \Phi(z_1) = \sum_{k=0}^{\infty}
\frac{\Phi^{(k)}(z_2)}{k!} \frac{\Gamma(k-1+\varepsilon)}
{\Gamma\left(k-\ell_1-\frac{1-\varepsilon}{2}-u\right)} \cdot
z_{12}^k =
$$
$$
= \Phi(z_2) \frac{\Gamma(-1+\varepsilon)}
{\Gamma\left(-\ell_1-\frac{1}{2}-u\right)}+ \Phi^{\prime}(z_2)
\frac{\Gamma(\varepsilon)}
{\Gamma\left(-\ell_1+\frac{1}{2}-u\right)}\cdot z_{12} +
O(\varepsilon^0)
$$
but we really have to extract singular part $\sim
\varepsilon^{-1}$ only
$$
\left[ \Phi(z_2)\left(-\ell_1-\textstyle{\frac{1}{2}}-u\right)+
\Phi^{\prime}(z_2)\cdot z_{21}\right]\cdot
\frac{\Gamma(-1+\varepsilon)}
{\Gamma\left(-\ell_1+\frac{1}{2}-u\right)}.
$$
Now we see that the simple pole cancels out the simple zero. The only
difference with the case $\ell_2 = 0$ is that we have to take into
account two simple poles of $\Gamma$-function instead of one
there. The action of $\R$-operator on components of $\Psi(z_1,z_2)
$ can be represented in the following form
$$
\phi(z_1)\rightarrow \biggl[
\phi(z_1)\left(-\ell_1-\textstyle{\frac{1}{2}}-u\right)+
\phi^{\prime}(z_1)\cdot z_{12}\biggl]\cdot
\frac{\Gamma\left(-\ell_1-\frac{1}{2}+u\right)}
{\Gamma\left(-\ell_1+\frac{1}{2}-u\right)}
$$
$$
z_2 \psi(z_1)\rightarrow
\biggl[z_1^2\psi^{\prime}(z_1)-2\ell_1z_1\psi(z_1)-
z_2\left(z_1\psi^{\prime}(z_1)+
\left(-\ell_1+\textstyle{\frac{1}{2}}+u\right)\psi(z_1)
\right)\biggl]\cdot
\frac{\Gamma\left(-\ell_1-\frac{1}{2}+u\right)}
{\Gamma\left(-\ell_1+\frac{1}{2}-u\right)}
$$
In the basis $\mathbf{e}_1 = -z_2 , \mathbf{e}_2 = 1$ we have for
$\ell_1 = \ell$ and $\ell_2 \to \frac{1}{2}$
$$
\R \left(u|\ell,\textstyle{\frac{1}{2}}\right) \mathbf{e}_1 =
-\frac{\Gamma\left(-\ell-\frac{1}{2}+u\right)}
{\Gamma\left(-\ell+\frac{1}{2}-u\right)}\cdot
\biggl[\mathbf{e}_1\left(z_1\dd_1-\ell+\textstyle{\frac{1}{2}}+u\right)
+\mathbf{e}_2\left(z_1^2\dd_1-2\ell z_1\right)\biggl]
$$
$$
\R \left(u|\ell,\textstyle{\frac{1}{2}}\right) \mathbf{e}_2 =
-\frac{\Gamma\left(-\ell-\frac{1}{2}+u\right)}
{\Gamma\left(-\ell+\frac{1}{2}-u\right)}\cdot
\biggl[\mathbf{e}_1\left(-\dd_1\right)
+\mathbf{e}_2\left(u+\textstyle{\frac{1}{2}}+\ell-z_1\dd_1\right)\biggl]
$$
and in operator form we finally obtain\footnote{In this formula we
use the notation $\mathrm{L}(u|\ell)$ which contains explicitly
spectral  and spin parameter and which should not be mixed up
with the other notation $\mathrm{L}(u_1,u_2)$}
\begin{equation} \label{R_1/2}
\R
\left(u|\ell,{\textstyle\frac{1}{2}}\right)\left|_{\mathbb{V}_{1}}\right.
= -\frac{\Gamma\left(-\ell-\frac{1}{2}+u\right)}
{\Gamma\left(-\ell+\frac{1}{2}-u\right)}\cdot
\mathrm{L}\left(u+{\textstyle\frac{1}{2}}|\ell\right) \, .
\end{equation}
Our calculations clearly demonstrate the cutting mechanism for the
degree of polynomials, i.e. why at special points $\ell =
\frac{n}{2}$ there appear finite-dimensional invariant subspaces
of the $\R$-operator.

Note that it is possible to use the reverse order of our two basic
operators in the expression for $\R$-operator~(\ref{Rfact}). Then
everything is finite and there is no need in any regularization.
The cutting mechanism is hidden in pure combinatorial
manipulations using the Pfaff-Saalsch\"utz formula so that we have
chosen the presented method of calculations as the most
illuminating.

And the last but not least: one of the reasons why we present here
these calculations is to show that only for the whole
$\R$-operator there exists the finite-dimensional invariant
subspaces for special values of spins but operators $\mathrm{R}^1$
or $\mathrm{R}^2$ separately map beyond  these subspaces. Keeping
this in mind it will be not very surprising to encounter this phenomenon on
the level of $\mathrm{Q}$-operators.

\subsection{Restriction of general $\R$-operator to finite-dimensional
representations}

In the previous subsection we have demonstrated by explicit
calculation how to restrict the general  $\R$-operator acting in
the space $\mathbb{U}_{-\ell_1}\otimes\mathbb{U}_{-\ell_2}$ to the
invariant subspace $\mathbb{U}_{-\ell_1}\otimes\mathbb{V}_{n}$.
For integer values of $2\ell_1 = n$ the reducibility appears in
the first tensor factor and the reduction to the
$(n+1)$-dimensional irreducible representation space can be done.
It is convenient to use the projection operators
\begin{equation}
\label{proj} \Pi^n_i z_i^k = z_i^k\ ,\  k\le n\ \ \ ; \ \ \
\Pi^n_i z_i^k = 0\ ,\ k > n.
\end{equation}
We shall write in boldface style the operators related to the
integer or half-integer spin case which action is restricted to
the finite-dimensional irreducible subspace
\begin{equation} \label{rest}
\mathbf{R}_{12}\left(u|{\textstyle\frac{n}{2}}, \ell_2\right) =
\R_{12} \left(u|{\textstyle\frac{n}{2}}, \ell_2\right) \Pi_1^n\ \
; \ \ \mathbf{R}_{12}\left(u|\ell_1,{\textstyle\frac{n}{2}}\right)
= \R_{12} \left(u|\ell_1,{\textstyle\frac{n}{2}}\right) \Pi_2^n
\end{equation}
The restriction in the second tensor factor is needed for
construction of transfer matrices with finite-dimensional
auxiliary space. In the previous section we have analyzed two
particular cases needed for the derivation of the Baxter equation.

\subsubsection{Operator $\mathbf{R}$}

Now we concentrate on the restriction in the first tensor factor
which corresponds to the restriction in quantum space and consider
the limit $2\ell_1 \to n ,\ n= 0 , 1 , 2 ,\cdots $ in the operator
\begin{equation}
\label{PR1} \mathrm{R}(u|\ell_1,\ell_2) = e^{-z_1 \dd_2}
\frac{\Gamma(z_2\partial_2 - 2\ell_1)}{\Gamma (z_2\partial_2 -
\ell_1-\ell_2 - u )} e^{z_1 \dd_2} \cdot e^{-z_2 \dd_1}
\frac{\Gamma(z_1\partial_1 - \ell_1-\ell_2+u)}{\Gamma
(z_1\partial_1 - 2\ell_1 )} e^{z_2 \dd_1}\,.
\end{equation}
We put $2\ell_1 = n -\varepsilon$ and analyze the limit
$\varepsilon \to 0$ which is governed by the behaviour of the
operator $\Gamma(z\partial - n + \varepsilon)$ and its inverse for
$\varepsilon \to 0$. The behaviour at small $\varepsilon$ becomes
more explicit after splitting operators into the contributions
acting on the monomials $z^k$ with $k\le n$ and $k>n$ by means of
the appropriate projectors~(\ref{proj}): $\Pi^n$ for monomials
with $k\le n$ and $1-\Pi^n$ for monomials with $k>n$ and using
Euler's reflection formula $ \Gamma(z)\,\Gamma(1-z) =
\frac{\pi}{\sin(\pi z)}$ for the contributions with projector
$\Pi^n$
$$
\Gamma(z_2\partial_2 - n+\varepsilon) = \Gamma( z_2\partial_2
-n+\varepsilon)\,(1-\Pi_2^n) + \frac{\pi}{\sin \pi
\varepsilon}\,\frac {(-1)^{z_2\partial_2+n}}{\Gamma
(1+n-\varepsilon -z_2\partial_2)}\, \Pi_2^n \,,
$$
$$
\frac{1}{\Gamma(z_1\partial_1 - n+\varepsilon)} = \frac{1}{\Gamma(
z_1\partial_1 -n+\varepsilon)}\,(1-\Pi_1^n) + \frac{\sin \pi
\varepsilon}{\pi}\, (-1)^{z_1\partial_1+n}\,\Gamma
(1+n-\varepsilon -z_1\partial_1)\, \Pi_1^n \,.
$$
The first contributions with projector $1-\Pi^n$ are nonsingular
for $\varepsilon\to 0$ and the second contribution clearly shows
that the operator $\Gamma(z_2\partial_2 - n+\varepsilon)$ diverges
on monomials $z_2^k, k \le n$ and the operator
$\Gamma^{-1}(z_1\partial_1 - n+\varepsilon)$ annihilates all
monomials $z_1^k, k \le n$ in the limit $\varepsilon \to 0$. We see
 that the nonsingular contribution with projector $1-\Pi^n$ does
not contribute  to a consistent restriction.  The
restriction to the subspace of monomials $z_1^k, k \le n$
leads to
the operator $\mathbf{R}_{12}\left(u|{\textstyle\frac{n}{2}},
\ell_2\right)$ (\ref{rest}).
Since
$e^{z_2 \dd_1}\,\Pi_1^n = \Pi_1^n\, e^{z_2
\dd_1}\, \Pi_1^n$ the projector $\Pi_1^n$ extracts the corresponding
contribution in the first factor in (\ref{PR1}),
$$
e^{-z_2 \dd_1} \frac{\Gamma(z_1\partial_1 -
\ell_1-\ell_2+u)}{\Gamma (z_1\partial_1 - 2\ell_1 )} e^{z_2
\dd_1}\Pi_1^n =
$$
$$
= \frac{(-1)^{n}\sin \pi \varepsilon}{\pi}\cdot e^{-z_2 \dd_1}\,
(-1)^{z_1\partial_1}\Gamma(z_1\partial_1 - \ell_1-\ell_2+u)\Gamma
(1+n-\varepsilon -z_1\partial_1)\, e^{z_2 \dd_1} \Pi_1^n\,.
$$
It holds for arbitrary $\varepsilon$, but vanishes in the limit
$\varepsilon \to 0$. Therefore one needs the
singular contribution $\sim\frac{1}{\varepsilon}$ from the second
operator only
$$
e^{-z_1 \dd_2} \frac{\Gamma(z_2\partial_2 - 2\ell_1)}{\Gamma
(z_2\partial_2 - \ell_1-\ell_2 - u )} e^{z_1 \dd_2} \to
\frac{(-1)^{n}\pi}{\sin \pi \varepsilon}\cdot e^{-z_1
\dd_2}\frac{(-1)^{z_2\partial_2}\Pi^n_2}{\Gamma(z_2\partial_2 -
\ell_1-\ell_2-u)\Gamma (1+n-\varepsilon -z_2\partial_2)} e^{z_1
\dd_2}
$$
Finally  one obtains the following explicit expression for
the restricted $\mathrm{R}$-operator
$$
\mathbf{R}_{12}\left(u|{\textstyle\frac{n}{2}}, \ell_2\right) =
\mathrm{P}_{12}\cdot e^{-z_1 \dd_2}\frac{(-1)^{z_2\partial_2}\,
}{\Gamma(z_2\partial_2 -\frac{n}{2}-\ell_2-u)\Gamma (1+n
-z_2\partial_2)} \, \Pi^n_2 \, e^{z_1 \dd_2} \cdot
$$
\be \label{Norm1}
\cdot e^{-z_2 \dd_1}\, (-1)^{z_1\partial_1}\Gamma(z_1\partial_1
-\textstyle{\frac{n}{2}}-\ell_2+u)\,\Gamma (1+n -z_1\partial_1)\,
e^{z_2 \dd_1} \Pi_1^n\,.
\ee
 Note that finite-dimensional subspace is preserved
under the action of the operator
$\mathbf{R}_{12}\left(u|{\textstyle\frac{n}{2}}, \ell_2\right)$
because projector $\Pi_1^n$ appears not only on the right but on
the left too
\be \label{Sproperty}
\mathrm{P}_{12}\, e^{-z_1\partial_2}\, \Pi_2^n = \mathrm{P}_{12}\,
\Pi_2^n \, e^{-z_1\partial_2}\, \Pi_2^n = \Pi_1^n\,
\mathrm{P}_{12}\,e^{-z_1\partial_2}\, \Pi_2^n\,.
\ee

\subsubsection{Operators $\mathbf{R}^1$,
$\mathbf{R}^2$ and $\mathbf{S}$} \label{R1R2Sfindim}

Now we  consider special reductions of (\ref{Norm1})
by specifying the values of the parameters $v_1, v_2$.
We rewrite the expression for
$\mathbf{R}_{12}\left(u-v|{\textstyle\frac{n}{2}}, \ell_2\right)$
using the parametrization
\be \label{paramfindim}
u_1 = u - {\textstyle\frac{n}{2}} -1, \ u_2 = u +
{\textstyle\frac{n}{2}}\ ;\ v_1 = v - \ell_2 -1, \ v_2 = v +
\ell_2\,,
\ee
and study  the limits $v_1\to u_1$ or (and) $v_2\to
u_2$.
We have $\mathbf{R}_{12}(u_1 , u_2|v_1,v_2) =
\mathbf{R}_{12}(u-v|{\textstyle\frac{n}{2}}, \ell_2)$ so that
\begin{equation}\label{Norm2} \mathbf{R}_{12}(u_1 , u_2|v_1,v_2)
= \mathrm{P}_{12}\cdot e^{-z_1
\dd_2}\,\frac{(-1)^{z_2\partial_2}}{\Gamma(z_2\partial_2
+v_1-u_2+1)\Gamma (u_2-u_1-z_2\partial_2)}\, \Pi^n_2\, e^{z_1
\dd_2}\cdot
\end{equation}
$$
\cdot e^{-z_2 \dd_1}\, (-1)^{z_1\partial_1}\, \Gamma(z_1\partial_1
+u_1-v_2+1)\Gamma (u_2-u_1-z_1\partial_1)\, e^{z_2 \dd_1}\,
\Pi_1^n\,.
$$

\begin{itemize}

\item $v_2\to u_2$

First we consider the limit of the operator $\mathbf{R}_{12}(u_1 ,
u_2|v_1,v_2)$ when $v_2\to u_2$. For this we put $v_2 = u_2
-\delta$ and using Euler's reflection formula derive the leading
contribution in the limit $\delta\to 0$
$$
(-1)^{z_1\partial_1}\, \Gamma(z_1\partial_1
+u_1-u_2+\delta+1)\Gamma(u_2-u_1-z_1\partial_1) \to
\frac{(-1)^{u_2-u_1-1}}{\delta}
$$
so that one obtains
$$
\mathbf{R}_{12}(u_1 , u_2|v_1,u_2-\delta) \to
\delta^{-1}\cdot\mathbf{R}^1_{12}(u_1 |v_1,u_2)\,,
$$
where we defined some operator which is the relative of the
operator $\R^1_{12}$ ( \ref{R1R2}) in the case of
finite-dimensional representations
\begin{equation}\label{R1findim}
\mathbf{R}^1_{12}(u_1 |v_1,u_2) \equiv \mathrm{P}_{12}\cdot e^{-z_1
\dd_2} \frac{(-1)^{z_2\partial_2+u_2-u_1-1}}{\Gamma(z_2\partial_2
+v_1-u_2+1)\Gamma (u_2-u_1-z_2\partial_2)}\,\Pi^n_2\, e^{z_1
\dd_2} \cdot \Pi_1^n
\end{equation}

\item $v_1\to u_1$

Next we consider the limit of the operator $\mathbf{R}_{12}(u_1 ,
u_2|v_1,v_2)$ for $v_1\to u_1$. For this we put $v_1 = u_1
+ \delta$ and using Euler's reflection formula derive the leading
contribution in the limit $\delta\to 0$
$$
\frac{(-1)^{z_2\partial_2}}{\Gamma(z_2\partial_2
+u_1-\delta-u_2+1)\Gamma (u_2-u_1-z_2\partial_2)} \to
(-1)^{u_2-u_1-1}\,\delta
$$
so that one obtains
$$
\mathbf{R}_{12}(u_1 , u_2|u_1+\delta,v_2) \to
\delta\cdot\mathbf{R}^2_{12}(u_1, u_2 |v_2)\,,
$$
where we defined the  operator which is the relative of the
operator $\R^2_{12}$ (\ref{R1R2}) in the case of
finite-dimensional representations
\begin{equation}\label{R2findim}
\mathbf{R}^2_{12}(u_1,u_2 |v_2)  \equiv \mathrm{P}_{12}\cdot e^{-z_1
\dd_2}\, \Pi^n_2\, e^{z_1 \dd_2}\cdot
\end{equation}
$$
\cdot e^{-z_2 \dd_1}\,
(-1)^{z_1\partial_1+u_2-u_1-1}\,\Gamma(z_1\partial_1 +u_1-v_2+1)\,
\Gamma (u_2-u_1-z_1\partial_1)\, e^{z_2 \dd_1}\, \Pi_1^n
$$
In comparison with (\ref{R1findim}) the last operator is more complicated.

\item $v_1\to u_1$ and $v_2\to u_2$

Finally we consider the limit of the operator $\mathbf{R}_{12}(u_1 ,
u_2|v_1,v_2)$ when simultaneously $v_1\to u_1$ and $v_2\to u_2$.
We put $v_1 = u_1+\delta$ and $v_2 = u_2-\delta$ where $\delta \to 0$.
As before we use Euler's reflection formula and obtain
\begin{equation}\label{Rfindim}
\mathbf{R}_{12}(u_1 , u_2|u_1 + \delta ,u_2 - \delta ) \to \mathbf{S}_{12} \equiv
\mathrm{P}_{12}\cdot e^{-z_1 \dd_2}\,\Pi^n_2\, e^{z_1 \dd_2} \cdot
\Pi_1^n 
\end{equation}
 Contrary to the infinite-dimensional representation case,
$\mathbf{R}_{12}(u_1,u_2|v_1,v_2)$ has
the  limit $\mathbf{S}_{12}$ that is not equal to transposition.
This result clearly demonstrates
that in order to construct operators for
finite-dimensional representation in the quantum space
one has to perform carefully the appropriate limiting procedures.

\end{itemize}

\subsubsection{Connection between two sets of operators} \label{R1R2Sfindim'}

Now we are going to establish the connection between the operators
$\mathbf{R}^1_{12}(u_1 |v_1,u_2)$ and $\mathbf{R}^2_{12}(u_1, u_2
|v_2)$ introduced in (\ref{R1findim}), (\ref{R2findim}) at special
relations on parameters (\ref{paramfindim}) and appropriate limits
of operators $\R^1_{12}(u_1 |v_1,u_2)$, $\R^2_{12}(u_1 u_2|v_2)$
(\ref{R1R2}).

We shift by $\varepsilon$ the spin in quantum space: $\ell =
\frac{n}{2} \to \frac{n}{2}-\frac{\varepsilon}{2}$ and
correspondingly $u_1 = u - \frac{n}{2} -1\to
u_1+\frac{\varepsilon}{2}$ and $u_2 = u +\frac{n}{2}\to u_2
-\frac{\varepsilon}{2}$ and consider the limit $\varepsilon\to 0$.

We start with $\R^1_{12}(u_1+{\textstyle\frac{\varepsilon}{2}}
|v_1,u_2-{\textstyle\frac{\varepsilon}{2}})$. The leading
contribution in the limit $\varepsilon\to 0$ has the form
$$
\R^1_{12}(u_1+{\textstyle\frac{\varepsilon}{2}}
|v_1,u_2-{\textstyle\frac{\varepsilon}{2}} ) \to
 \frac{1}{\varepsilon} \cdot \mathrm{P}_{12}\cdot e^{-z_1 \dd_2}
\frac{(-1)^{z_2\partial_2+u_2-u_1-1}}{\Gamma(z_2\partial_2
+v_1-u_2+1)\Gamma (u_2-u_1-z_2\partial_2)}\,\Pi^n_2\, e^{z_1
\dd_2}\,.
$$
The proof of this formula goes parallel to the above derivation of
the expression for the operator
$\mathbf{R}_{12}\left(u|{\textstyle\frac{n}{2}}, \ell_2\right)$:
we split $\R^1_{12}(u_1+{\textstyle\frac{\varepsilon}{2}}|v_1,u_2-
{\textstyle\frac{\varepsilon}{2}})$ using projectors $\Pi_2^n$ and
$1-\Pi_2^n$, multiply by $\varepsilon$ and obtain that in the
limit $\varepsilon\to 0$ only the singular contribution with
$\Pi_2^n$ survives.

The comparison with (\ref{R1findim}) gives another way for the
calculation of $\mathbf{R}^1_{12}(u_1 |v_1,u_2)$ starting
directly from his ancestor $\R^1_{12}(u_1 |v_1,u_2)$
\begin{equation}\label{relation1}
\mathbf{R}^1_{12}(u_1 |v_1,u_2) =  \lim_{\varepsilon\to 0}
\,\varepsilon\cdot \R^1_{12}(u_1+\textstyle{\frac{\varepsilon}{2}}
|v_1,u_2-\textstyle{\frac{\varepsilon}{2}})\,\Pi_1^n
\end{equation}
 The last formula means that
 $\R^1_{12}(u_1 |v_1,u_2)$ after renormalization
in the limit $\ell_1 \to \frac{n}{2}$ does not map beyond
the subspace $\mathbb{V}_n\otimes\mathbb{U}_{-\ell_2}$.
Moreover from (\ref{Sproperty}) one can see that it maps
$\mathbb{U}_{-\frac{n}{2}}\otimes\mathbb{U}_{-\ell_2}$ to
$\mathbb{V}_n\otimes\mathbb{U}_{-\ell_2}$.

Now we turn to the next pair of operators. This time the relation
is more complicated: $\mathbf{R}^2_{12}(u_1,u_2 |v_2)$ can be
obtained from the operator $\R^2_{12}(u_1,u_2|v_2)$ but there
exists the nontrivial first factor \be \label{S12} \S_{12} \equiv
\mathrm{P}_{12}\,e^{-z_1 \dd_2}\, \Pi^n_2\, e^{z_1 \dd_2} \ee so
that the final formulae of connection is
\begin{equation}\label{relation2}
\mathbf{R}^2_{12}(u_1,u_2 |v_2) = \S_{12}\cdot
\mathrm{P}_{12}\cdot \lim_{\varepsilon\to 0}
\,\frac{1}{\varepsilon}\cdot
\R^2_{12}(u_1+{\textstyle\frac{\varepsilon}{2}} ,
u_2-{\textstyle\frac{\varepsilon}{2}}|v_2)\,\Pi_1^n
\end{equation}
The proof is very similar to the proof of~(\ref{relation1}). This
formula means that the renormalized operator $\R^2_{12}(u_1 ,
u_2|v_2)$ in the limit $\ell_1 \to \frac{n}{2}$ maps beyond the
subspace $\mathbb{V}_n\otimes\mathbb{U}_{-\ell_2}$ and therefore
the correcting operator $\S_{12}$ is indispensable. Indeed
$\S_{12}$ maps
$\mathbb{U}_{-\frac{n}{2}}\otimes\mathbb{U}_{-\ell_2}$ to
$\mathbb{V}_n\otimes\mathbb{U}_{-\ell_2}$ as one can see from
(\ref{Sproperty}).

Let us point out the connection between the double reduction of
$\mathbf{R}_{12}(u_1 ,u_2|v_1,v_2)$ (\ref{Rfindim}) and operator
$\S_{12}$ (\ref{S12})
$$
\mathbf{S}_{12} = \S_{12} \cdot \Pi_1^n
$$

\subsection{The general transfer matrices and $\mathbf{Q}$-operators}

After the necessary preparations in the previous section we proceed to
the construction of the general transfer matrix (\ref{T}) from the
new building blocks $\mathbf{R}_{k0}$. Here we meet a certain
difficulty: the trace over the infinite-dimensional auxiliary space
$\mathbb{C}[z_0]$ diverges. Consequently we have to introduce some
kind of regularization. We shall use the following regularization
$$
\tr_{_0}\mathbf{B} \longrightarrow \tr_{_0} q^{z_0 \dd_0} \mathbf{B}
\quad , \quad |q|<1
$$
which corresponds to quasiperiodic boundary conditions of the spin
chain. A shortcoming of this regularization is a violation of
$s\ell_2$-symmetry in the traces and correspondingly in the
transfer matrices. Thus we define the general transfer matrix for
integer or half-integer values of $\ell$ and $|q|<1$ as
\be \label{Tfin}
\mathbf{T}_{s}(u) = \tr_0  \, q^{z_0 \dd_0} \,
\mathbf{R}_{10}\left(u|{\textstyle\frac{n}{2}}, s\right)
\mathbf{R}_{20}\left(u|{\textstyle\frac{n}{2}}, s\right) \cdots
\mathbf{R}_{N0}\left(u|{\textstyle\frac{n}{2}}, s\right) \,. \ee
This operator is well defined on the finite-dimensional quantum space of
the chain.


Now we consider the factorization of the general transfer
matrix (\ref{Tfin}). We start as before with the three term relation
(\ref{rf2'}) and restrict it at site $k$  to
$\mathbb{V}^n\otimes \mathbb{C}[z_0]\otimes \mathbb{C}[z_{0'}]$ for
$\ell=\frac{n}{2}$
$$
\R^2_{00^{\prime}}(v_1,v_2|w_2)
\mathbf{R}_{k0^{\prime}}(u_{1},u_2|w_1,w_2)
\mathbf{R}_{k0}(u_{1},u_2|v_1,v_{2}) =
$$
\be \label{rf2''} =\mathbf{R}_{k0}(u_{1},u_2|w_1,v_{2})
\mathbf{R}_{k0^{\prime}}(u_1,u_{2}|v_1,w_{2})
\R^2_{00^{\prime}}(v_{1},v_{2}|w_2)\, . \ee
Then we specify
parameters as $w_1 = u_1+\delta\,,\,w_2 =
u_2-\delta$ and obtain in the limit $\delta\to 0$
keeping in mind the results of Section~\ref{R1R2Sfindim}
$$
\R^2_{00^{\prime}}(v_1,v_2|v_2) \mathbf{S}_{k0^{\prime}}
\mathbf{R}_{k0}(u_{1},u_2|v_1,v_{2}) =
\mathbf{R}^2_{k0}(u_{1},u_2|v_{2})
\mathbf{R}^1_{k0^{\prime}}(u_1|v_1,u_{2})
\R^2_{00^{\prime}}(v_{1},v_{2}|u_2)\, .
$$
This local relation leads in the standard way to the factorization
relation for the corresponding regularized transfer matrices if we
take into account that $ [ \, q^{z_0 \dd_0 +
z_{0^{\prime}}\dd_{0^{\prime}}} \, , \, \R_{0 0^{\prime}} \, ] = 0
\,$

$$ \tr_{0'}
\left[ \, q^{z_{0'}\dd_{0'}} \, \mathbf{S}_{10'} \, \mathbf{S}_{20'} \cdots \mathbf{S}_{N0'} \,
\right]\cdot
\tr_{0} \left[\, q^{z_{0}\dd_{0}} \,\mathbf{R}_{10}(u_{1},u_2|v_1,v_{2})\cdots
\mathbf{R}_{N0}(u_{1},u_2|v_1,v_{2})\right] =
$$
$$ =\tr_{0} \left[\, q^{z_{0}\dd_{0}}
\mathbf{R}^2_{10}(u_{1},u_2|v_{2})\cdots
\mathbf{R}^2_{N0}(u_{1},u_2|v_{2})\right]\cdot\tr_{0'} \left[\,
q^{z_{0'}\dd_{0'}} \,\mathbf{R}^1_{10'}(u_{1}|v_1,u_{2})\cdots
\mathbf{R}^1_{N0'}(u_{1}|v_1,u_{2})\right]
$$
The second case of factorization can be obtained in a similar
way. After introduction of the notation for the
transfer matrices
$$ \mathbf{Q}_1(u-v_1) = \tr_{0}
\, q^{z_0 \dd_0} \, \mathbf{R}^1_{10}(u_1|v_1,u_{2})\cdots
\mathbf{R}^1_{N0}(u_1|v_1,u_{2})\,,
$$
$$
\mathbf{Q}_2(u-v_2) = \tr_{0} \, q^{z_0 \dd_0} \,
\mathbf{R}^2_{10}(u_1,u_2|v_{2})\cdots
\mathbf{R}^2_{N0}(u_1,u_2|v_{2})\,,
$$
\be \label{Sbold} \mathbf{S} = \tr_{0} \, q^{z_0\dd_0} \,
\mathbf{S}_{10} \, \mathbf{S}_{20} \cdots \mathbf{S}_{N0} \,,\ee
we can rewrite factorization relations in a compact form \be
\label{T->Q2Q1Fin} \mathbf{S}\,\mathbf{T}_{s} (u) = \mathbf{Q}_{2}
(u - s) \, \mathbf{Q}_{1} (u + s + 1) = \mathbf{Q}_{1} (u + s +
1)\, \mathbf{Q}_{2} (u - s) \,. \ee
This construction is analogous to the
construction from the first part, where $\ell$ was a generic
complex number. The proof of the commutativity \be
\label{commutQiQkFin} [ \, \mathbf{T}_{s}(u) , \mathbf{Q}_k (v) \,
] = 0 \ \ ;\ \ [ \, \mathbf{Q}_i (u) , \mathbf{Q}_k (v)\, ] =  0 \
\ ;\ \ [ \, \mathbf{S} \, , \, \mathbf{Q}_{k}(u) \, ]  = 0 \ \ ;\
\ [  \, \mathbf{S} \, ,\, \mathbf{T}_{s}(u)\,] = 0\,. \ee uses the
general Yang-Baxter equation and also goes parallel the corresponding
derivation given in the first part.

\subsection{Connection between  $\mathrm{Q}$-operators
of compact and generic spin and the Baxter equations}

In Section~\ref{R1R2Sfindim'} we have established relations
between  the two sets of $\mathrm{R}$-operators. Now we formulate
the corresponding relations for  the transfer matrices. Let us
consider the relation between the operators $\mathbf{Q}_{1}(u)$
and $\mathbf{Q}_{2}(u)$ and the limits at $\ell\to\frac{n}{2}$ of
the regularized Baxter $\mathrm{Q}$-operators
$\mathrm{Q}_{1}(u|q)$ and $\mathrm{Q}_{2}(u|q)$~(\ref{regQ}),
which are trivial modifications of operators $\mathrm{Q}_{1,2}(u)$
being constructed in sect.\ref{global}.

The connection between operators $\mathbf{Q}_{1}(u)$ and
$\mathrm{Q}_1(u|q)$ can be established directly using the
formula~(\ref{relation1}): we put the spin $\ell =
\frac{n}{2}-\frac{\varepsilon}{2}$ in quantum space and in the
limit $\varepsilon\to 0$ obtain
\begin{equation}\label{bfQ1}
\mathbf{Q}_1(u-v_1) = \lim_{\varepsilon\to 0} \varepsilon^N
\cdot\mathrm{Q}_{1}(u-v_1|q)\bigl|_{\ell =
\frac{n-\varepsilon}{2}}\,\cdot\, \Pi^n\,
\end{equation}
where $\Pi^n \equiv
\Pi^n_1\,\Pi^n_2\cdots\Pi^n_{N}$.

In our notations the boldface style in $\mathbf{Q}_1(u-v_1)$ means
that the parameter of the spin in this operator is half-integer
$\ell=\frac{n}{2}$ and the operator is restricted to the
appropriate finite-dimensional subspace by the projector $\Pi^n$.
But it is easy to see that there exists the extension of operator
$\mathbf{Q}_1(u-v_1)$ to the whole space of polynomials or in
other words there exists the limit
\begin{equation}\label{Q1ext}
\mathrm{Q}_1(u-v_1) = \lim_{\varepsilon\to 0} \varepsilon^N
\cdot\mathrm{Q}_{1}(u-v_1|q)\bigl|_{\ell =
\frac{n-\varepsilon}{2}}\,.
\end{equation}
To avoid making copies of notations we denote this extended
operator by the same notations as in the case of generic spin. We
hope it will not lead to misunderstanding and it will be clear
from the context which operator is used. The operator
$\mathrm{Q}_1(u-v_1)$ inherits properties from its local building
blocks. In particular, due to~(\ref{Sproperty}) this operator does
not map beyond the finite-dimensional quantum space and moreover
it maps the whole space of polynomials to the finite-dimensional
quantum space.

To derive the second relation we rely on the factorization
(\ref{factorII}) for arbitrary spin in quantum space. Now we need
a simple modification of this formula because of the
regularization
\begin{equation}\label{factorIIq}
\tr_{0^{\prime}} \left[ q^{z_{0^{\prime}}\partial_{0^{\prime}}}\,
\mathrm{P}_{10^{\prime}}\cdots\mathrm{P}_{N0^{\prime}}\right]\cdot
\tr_{0} \left[q^{z_{0}\partial_{0}}\,
\R_{10}(u_{1},u_2|v_1,v_{2})\cdots
\R_{N0}(u_{1},u_2|v_1,v_{2})\right] =
\end{equation}
$$
= \tr_{0} \left[q^{z_{0}\partial_{0}}\,
\R^{1}_{10}(u_{1}|v_1,u_{2})\cdots
\R^{1}_{N0}(u_{1}|v_1,u_{2})\right]\cdot \tr_{0^{\prime}} \left[
q^{z_{0^{\prime}}\partial_{0^{\prime}}}\,
\R^2_{10^{\prime}}(u_1,u_{2}|v_{2})\cdots
\R^2_{N0^{\prime}}(u_1,u_{2}|v_{2})\right].
$$
Next we multiply from the right by the  projector $\Pi^n =
\Pi^n_1\,\Pi^n_2\cdots\Pi^n_{N}$ which results in the restriction to
the appropriate subspace of the quantum space and put $2\ell = n
-\varepsilon$ or equivalently $u_1-u_2+1 = - n +\varepsilon$. It
remains to do the limit $\varepsilon \to 0$. In the left hand
side one obtains the operator
$$
\mathrm{P}\,q^{z_1\dd_1}\cdot \tr_{0}
\left[q^{z_{0}\partial_{0}}\,
\mathbf{R}_{10}(u_{1},u_2|v_1,v_{2})\cdots
\mathbf{R}_{N0}(u_{1},u_2|v_1,v_{2})\right]
$$
where we used  $\tr_{0^{\prime}}\left[ \, q^{z_{0'}
\dd_{0'}} \, \mathrm{P}_{10^{\prime}}\cdots
\mathrm{P}_{N0^{\prime}}\right] = \mathrm{P}\,q^{z_1\dd_1}$. In
the right hand side we have to do some rearrangement of
$\varepsilon$ to be sure that a finite limit exists for both
operators in the product
$$
\mathrm{Q}_{1}(u-v_1|q)\cdot\mathrm{Q}_{2}(u-v_2|q)\,\Pi^n =
\varepsilon^N\,\mathrm{Q}_{1}(u-v_1|q)\cdot\frac{1}{\varepsilon^N}
\,\mathrm{Q}_{2}(u-v_2|q)\,\Pi^n
$$
Note that in the limit $\varepsilon\to 0$ the first operator
$\varepsilon^N\,\mathrm{Q}_{1}(u-v_1|q)$ gives operator
$\mathrm{Q}_{1}(u-v_1)$~(\ref{Q1ext}),  the extended version of
the operator $\mathbf{Q}_1(u-v_1)$. It is easy to see that the
expression $\varepsilon^{-N}
\cdot\mathrm{Q}_{2}(u-v_2|q)\cdot\Pi^n$ also gives a well defined
operator in the limit $\varepsilon\to 0$. We shall study this
operator in detail in next subsection. Note, that contrary to the
previous case of operator $\mathrm{Q}_1(u)$ the extension of the
operator $\lim_{\varepsilon\to 0}\,\varepsilon^{-N}
\cdot\mathrm{Q}_{2}(u-v_2|q)\,\Pi^n$ to whole space does not
exist. The reason is that for finite $\varepsilon$ the result of
the action of $\mathrm{Q}_{2}$ on
 vectors of the subspace extracted by the projector $\Pi^n$ is
$\sim\varepsilon^N$
but $\sim 1 $ or lower powers of $\varepsilon$ for action on
the complementary subspace extracted by projector $1-\Pi^n$.
Due to this fact after
multiplication by $\varepsilon^{-N}$ one obtains finite results
in the first case and divergences in the second case.

We obtain the following relation
$$
\mathrm{P}\,q^{z_1\dd_1}\cdot \tr_{0}
\left[q^{z_{0}\partial_{0}}\,
\mathbf{R}_{10}(u_{1},u_2|v_1,v_{2})\cdots
\mathbf{R}_{N0}(u_{1},u_2|v_1,v_{2})\right] =
\mathrm{Q}_{1}(u-v_1)\cdot\lim_{\varepsilon\to 0}\,
\varepsilon^{-N} \,\mathrm{Q}_{2}(u-v_2|q)\,\Pi^n\,,
$$
Then we have to  specify $v_1 = u_1+\delta$ and take the limit
$\delta\to 0$. Finally we obtain \be \label{bfQ2}
\mathrm{P}\,q^{z_1\dd_1}\cdot\mathbf{Q}_2(u-v_2) = \mathrm{S}\cdot
\lim_{\varepsilon\to 0}\, \varepsilon^{-N}\,
\mathrm{Q}_{2}(u-v_2|q)\cdot\Pi^n\,, \ee where the new operator
$\mathrm{S}$ is the special limit of the operator $\mathrm{Q}_1$
(\ref{Q1ext}) and can be represented as a transfer matrix
constructed from the operators $\S_{k0}$ (\ref{S12}) \be
\label{SS} \mathrm{S} = \lim_{\delta\to 0}\ \delta^{-N}\,
\mathrm{Q}_{1}(\textstyle{\frac{n}{2}}+1-\delta) = \tr_{0} \,
q^{z_0\dd_0} \, \S_{10} \, \S_{20} \cdots \S_{N0} \,. \ee Like
$\mathrm{Q}_1$ this operator $\mathrm{S}$ maps the whole space of
polynomials to the finite-dimensional subspace. We see that for
the operator $\mathbf{Q}_2(u)$ the connection is more complicated:
it coincides up to normalization with the product of the nontrivial
operator $q^{-z_1\dd_1}\,\mathrm{P}^{-1}\, \mathrm{S}$ and the
restriction of $\mathrm{Q}_{2}(u|q)$ to the invariant subspace
appearing for $\ell = \frac{n}{2}$ . (\ref{bfQ2}) is a global
analogue of the local formula (\ref{relation2}). Above we have
seen on particular example (\ref{example}) that the renormalized
Baxter operator $\mathrm{Q}_2(u)$ maps beyond the
finite-dimensional subspace. The same is true for
$\lim_{\varepsilon\to 0}\, \varepsilon^{-N}\,
\mathrm{Q}_{2}(u|q)\cdot\Pi^n$. Now we see that the problem of
finite-dimensional representations is resolved by means of the
special operator $q^{-z_1\dd_1}\,\mathrm{P}^{-1}\, \mathrm{S}$.
Its role is to reverse the mapping back to the finite-dimensional
subspace after it
 went beyond by  the action of the limit of $\mathrm{Q}_{2}(u|q)$
 producing the correct Baxter operator
$\mathbf{Q}_2(u)$ on the finite-dimensional quantum space.

The relations (\ref{bfQ1}) and (\ref{bfQ2}) allow to obtain
explicit compact formulae for Baxter operators
$\mathbf{Q}_{1,2}(u)$. We postpone this derivation to the Section
\ref{explicit'}.

Now we turn to Baxter relations for $\mathbf{Q}_{1}(u)$ and $\mathbf{Q}_{2}(u)$.
Both proofs of Baxter relation presented above in sections
\ref{Qbyhand} and \ref{global} can be transferred step by step to
the case of half-integer $\ell$. However there is a simpler way
to establish such relations since we know
the relations (\ref{bfQ1}) and (\ref{bfQ2}) between
Baxter operators for finite and infinite-dimensional quantum spaces.
We follow this way below.

The introduction of the regularization leads to simple
modifications which are discussed in detail in Appendix C. Finally
the Baxter equations for regularized operators
$\mathrm{Q}_{1,2}(u|q)$ have the form \be \label{qBaxter1}
\mathrm{t}(u|q)\,\mathrm{Q}_1(u|q) =\mathrm{Q}_1(u+1|q)+
q\cdot(u_1 u_2)^N \cdot\mathrm{Q}_1(u-1|q)\,, \ee \be
\label{qBaxter2} \mathrm{t}(u|q) \, \mathrm{Q}_2(u|q) = q\cdot
\mathrm{Q}_2(u+1|q)+ (u_1 u_2)^N \cdot\mathrm{Q}_2(u-1|q)\,, \ee
where \be \label{regt} \mathrm{t}(u|q) = \tr
\begin{pmatrix} q & 0 \\ 0 & 1 \end{pmatrix}
\mathrm{L}_1(u)\mathrm{L}_2(u)\cdots \mathrm{L}_N(u)\,. \ee

In order to derive Baxter equation for $\mathbf{Q}_{1}$ we have to
multiply the obtained equation (\ref{qBaxter1}) by the projector
$\Pi^n$ from the right and specify the  spin parameter
$\ell=\frac{n}{2}$
\begin{equation}\label{finBaxter1}
\mathrm{t}(u|q)\mathbf{Q}_1(u) =  \mathbf{Q}_1(u+1)+ q\cdot(u_1
u_2)^N\cdot\mathbf{Q}_1(u-1)\,.
\end{equation}
The Baxter equation for operator $\mathbf{Q}_{2}$:
\begin{equation}\label{finBaxter2}
\mathrm{t}(u|q) \, \mathbf{Q}_2(u) = q \cdot \mathbf{Q}_2(u+1)+
(u_1 u_2)^N \cdot \mathbf{Q}_2(u-1)\,
\end{equation}
is derived in a similar way from (\ref{qBaxter2}) but there is one
additional step - the multiplication by the operator
$q^{-z_1\dd_1}\,\mathrm{P}^{-1}\, \mathrm{S}$ from the left. This
operator does not depend on spectral parameter so that this step
does not change the equation.

Thus we see that $\mathbf{Q}_{1}$ and $\mathbf{Q}_{2}$ possess all
expected properties of Baxter operators. Moreover because they
are obtained at special values of parameters from the general transfer
matrix (\ref{Tfin}) they map finite-dimensional space into itself.
However they are not $s\ell_2$-invariant because of
$q$-regularization.

\subsection{Explicit  action on polynomials}
\label{explicit'}

Now we are going to consider explicit formulae for the action of
the constructed operators $\mathbf{Q}_k(u)$ on polynomials. For this
purpose we shall use the connection with the operators
$\mathrm{Q}_k(u|q)$ on infinite quantum space.

It is easy to repeat step by step all derivations from the
section~\ref{explicit} and obtain formulae for the operators
$\mathrm{Q}_k(u|q)$ which simply mimic formulae for the operators
$\mathrm{Q}_k(u)$ with needed minimal modifications due to
q-regularization. The expression for the action of the operator
$\mathrm{Q}_2(u|q)$ on polynomials is the following
\begin{equation}\label{bfQ2q}
\mathrm{Q}_{2}(u|q)\,\Psi(\vec{z}) =
\left.\mathrm{P} \cdot \mathrm{R}_2(\lambda_1\partial_{\lambda_1})\cdots
\mathrm{R}_2(\lambda_N\partial_{\lambda_N})\right|_{\lambda=1}\cdot
\Psi(\Lambda_q\vec{z})\,,
\end{equation}
where
$$
\Lambda_q = \begin{pmatrix}
q \lambda_1 & 1-\lambda_1 & 0 & 0 & \hdotsfor{2} & 0 \\
0 & \lambda_2 & 1-\lambda_2 & 0 & \hdotsfor{2} & 0 \\
0 & 0 & \lambda_3 & 1-\lambda_3 & \hdotsfor{2} & 0 \\
\hdotsfor{7}  \\
0 & 0 & \hdotsfor{3} & \lambda_{N-1} & 1-\lambda_{N-1} \\
1-\lambda_{N} & 0 & 0 & 0 & \cdots & 0 & \lambda_{N}
\end{pmatrix} \ \ ;\ \ \vec{z} = \begin{pmatrix}
z_1\\ z_2\\ z_3\\ \hdotsfor{1} \\ \hdotsfor{1} \\ z_{N}
\end{pmatrix}
$$
and the expression for the action of the operator
$\mathrm{Q}_1(u|q)$ on polynomials is very similar to the
corresponding formula for $\mathrm{Q}_2(u|q)$
\begin{equation}\label{bfQ1q}
\mathrm{Q}_{1}(u|q)\,\Psi(\vec{z}) =
\left.\mathrm{R}_1(\lambda_1\partial_{\lambda_1})\cdots
\mathrm{R}_1(\lambda_N\partial_{\lambda_N})\right|_{\lambda=1}\cdot
\frac{1}{ 1 - q\bar{\lambda}_1 \cdots \bar{\lambda}_N } \cdot
\Psi(\Lambda_q^{\prime -1}\, \vec{z} \, ) \,,
\end{equation}
where $\bar\lambda \equiv 1- \lambda$ and
$$
\Lambda_q^{\prime} = \begin{pmatrix}
1-\frac{1}{\lambda_1} & \frac{1}{\lambda_1} & 0 & 0 & \hdotsfor{2} & 0 \\
0 & 1-\frac{1}{\lambda_2} & \frac{1}{\lambda_2} & 0 & \hdotsfor{2} & 0 \\
0 & 0 & 1-\frac{1}{\lambda_3} & \frac{1}{\lambda_3} & \hdotsfor{2} & 0 \\
\hdotsfor{7}  \\
0 & 0 & \hdotsfor{3} & 1-\frac{1}{\lambda_{N-1}} & \frac{1}{\lambda_{N-1}} \\
\frac{1}{q \lambda_{N}} & 0 & 0 & 0 & \cdots & 0 &
1-\frac{1}{\lambda_{N}}
\end{pmatrix} \,.
$$
Formulae~(\ref{bfQ2q}) and~(\ref{bfQ1q}) are the starting points for
the derivation of various representations for the
$\mathrm{Q}$-operators.  Integral formulae similar
to~(\ref{Q-1}) and~(\ref{Q-2}) are obtained by evident changes in
matrices $\Lambda$ so that we shall not repeat all these formulae
but instead concentrate on the derivation of the useful
representation for operator $\mathrm{Q}_1$ in the case of
half-integer spin.

We use the following integral representation for all
operators $\mathrm{R}_1(\lambda_k\dd_{\lambda_k})$
$$
\mathrm{R}_1(\lambda\dd_{\lambda})\left.\Phi(\lambda)\right|_{\lambda
= 1} = \frac{1} {\Gamma(1+\ell-u)}\cdot
\int^{1}_{0}\mathrm{d}\lambda (1-\lambda)^{\ell-u}
\lambda^{-2\ell-1}\Phi(\lambda) \,.
$$
Note that for $2\ell = n -\varepsilon$ we have the pole
$\sim\frac{1}{\varepsilon}$ due to divergence in the integral
arising from the pole $\lambda^{-n-1}$ in the integrand at
$\varepsilon = 0$. Because of the factor $\varepsilon^N$ in the
definition of the operator $\mathrm{Q}_1(u)$ (\ref{Q1ext})  we
have to calculate only the singular contribution $\sim
\frac{1}{\varepsilon}$ in each integral resulting in a significant
simplification of the calculation.
$$
\int^{1}_{0}\mathrm{d}\lambda (1-\lambda)^{\ell-u}
\lambda^{-2\ell-1}\Phi(\lambda)  \to  \frac{1}{\varepsilon}\,
\left.\frac{\partial_{\lambda}^{n}}{n!}\,
(1-\lambda)^{\frac{n}{2}-u}\,\Phi(\lambda) \right|_{\lambda = 0}
$$
Finally we arrive at a compact formula for the action of the operator
$\mathrm{Q}_1(u)$
\begin{equation}\label{Q1findim}
\mathrm{Q}_{1}(u) \Psi(\vec{z} ) = \frac{1}
{\Gamma^N(1+\frac{n}{2}-u)\,n!^N}\cdot\dd^n_{\lambda_1}\cdots
\dd^n_{\lambda_N} \left. \frac{\left(\bar{\lambda}_1 \cdots
\bar{\lambda}_N\right)^{\frac{n}{2}-u} }{ 1 - q \bar{\lambda}_1
\cdots \bar{\lambda}_N } \cdot \Psi(\Lambda_{q}^{\prime -1}
\vec{z} \, ) \right|_{\lambda = 0} \,.
\end{equation}
This operator is defined on the whole infinite-dimensional
quantum space. In order to calculate the action
of $\mathbf{Q}_1(u)$ on arbitrary polynomial $\Psi(\vec{z} )$
 according to (\ref{bfQ1})
one has to pick up powers of $z_1, \cdots , z_N$ less or equal $n$ and
then apply
formula (\ref{Q1findim}) to the obtained polynomial.

At the point of degeneracy $u=\frac{n}{2}+1-\delta$ in the
appropriate limit $\delta\to 0$ the operator $\mathrm{Q}_{1}(u)$
reduces to the operator $\mathrm{S}$ (\ref{SS})
\begin{equation}\label{S'}
\mathrm{S} \Psi(\vec{z} ) =
\left(\frac{1}{n!}\right)^N\,\dd^n_{\lambda_1}\cdots
\dd^n_{\lambda_N} \left. \frac{1}{\bar{\lambda}_1 \cdots
\bar{\lambda}_N}\,\frac{1}{ 1 - q \bar{\lambda}_1 \cdots
\bar{\lambda}_N } \cdot \Psi(\Lambda_{q}^{\prime -1} \vec{z} \, )
\right|_{\lambda=0} \,
\end{equation}
This operator as well as $\mathrm{Q}_{1}(u)$ is defined on the
whole infinite-dimensional quantum space. It is evident how to
calculate $\mathbf{S}$ (\ref{Sbold}) on arbitrary polynomial
$\Psi(\vec{z}\, )$ because $\mathbf{S} = \mathrm{S}
\cdot \Pi^n$.

The expression (\ref{S'}) can be derived directly.
At first let us establish
how $\mathbb{S}_{k0}$ (\ref{S12}) acts on the function of two
arguments
$$
\mathbb{S}_{k0} \, \Phi(z_k,\,z_0) = \mathrm{P}_{k0} \, e^{-z_k
\dd_0} \, \Pi^n_0 \, \Phi(z_k,\,z_0+z_k)= \mathrm{P}_{k0}
\sum^{n}_{m=0} \frac{z^m_{0k}}{m!} \, \dd^m_{z} \,
\Phi(z_k,z)\bigl|_{z=z_k} =
$$
$$
= \sum^{n}_{m=0} \frac{\dd^m_{\lambda}}{m!} \, \Phi(z_0, z_0 +
\lambda z_{k0})\bigl|_{\lambda=0} = \sum^{n}_{m=0}
\frac{\dd^m_{\lambda}}{m!} \, \lambda^{z_{k0}\dd_k} \,
\Phi(z_0,z_k)\bigl|_{\lambda=0}\,.
$$
Thus we have
$$
\mathbb{S}_{k0} = \mathbf{e}_{n}(\dd_{\lambda})\cdot
\mathrm{P}_{k0}\cdot\lambda^{z_{0k}\dd_0}\bigl|_{\lambda = 0}
\quad ; \quad \mathbf{e}_{n}(\dd_{\lambda}) \equiv \sum^{n}_{m=0}
\frac{\dd^m_{\lambda}}{m!}\,,
$$
and consequently $\mathrm{S}$ (\ref{SS}) takes the form
$$ \mathrm{S} =
\mathbf{e}_{n}(\dd_{\lambda_1})\cdots
\mathbf{e}_{n}(\dd_{\lambda_N})
\tr_{\mathbb{V}_{0}}
\left.\left[\, q^{z_0\dd_0}\,
\mathrm{P}_{10}\lambda_1^{z_{01}\dd_0}\cdots
\mathrm{P}_{N0}\lambda_N^{z_{0N}\dd_0}
\,\right]
\right|_{\lambda=0}.
$$

The involved trace  is calculated by using (\ref{TraceExpl})
\be \label{Spolinom} \mathrm{S} \, \Psi(\vec{z}) =
\left. \mathbf{e}_{n}(\dd_{\lambda_1})\cdots
\mathbf{e}_{n}(\dd_{\lambda_N})\, \frac{1}{ 1 - q \bar{\lambda}_1
\cdots  \bar{\lambda}_N } \cdot \Psi(\Lambda_q^{\prime -1}\, \vec{z} \, )
\right|_{\lambda=0} \,
\ee
It remains to note that
$$
\left.\mathbf{e}_{n}(\dd_{\lambda}) \, \Psi\left(\lambda
\right)\right|_{\lambda=0} = \left. \frac{\dd_{\lambda}^n}{n!} \,
\frac{1}{1-\lambda} \,\Psi\left(\lambda\right)\right|_{\lambda=0}
$$
so that we are coming back to  formula~(\ref{S'}).

Finally we change normalization of Baxter operators
$\mathbf{Q}_1(u)$ and $\mathbf{Q}_2(u)$ in order to
make them to become polynomials in the spectral parameter $u$.
The explicit action of the  renormalized $\mathbf{Q}_1(u)$ (\ref{Q1findim})
which we denote $\mathbf{P}(u)$
has the form
\begin{equation}
\mathbf{P}(u) \Psi(\vec{z} ) = \dd^n_{\lambda_1}\cdots
\dd^n_{\lambda_N} \left. \frac{\left(\bar{\lambda}_1 \cdots
\bar{\lambda}_N\right)^{\frac{n}{2}-u} }{ 1 - q \bar{\lambda}_1
\cdots \bar{\lambda}_N } \cdot \Psi(\Lambda_{q}^{\prime -1}
\vec{z} \, ) \right|_{\lambda = 0} \,
\end{equation}
where $\Psi(\vec{z}\,)$ is polynomial from finite-dimensional invariant subspace.

The explicit  action of the renormalized $\mathbf{Q}_2(u)$ which
we denote $\mathbf{Q}(u)$ on the generating function of
finite-dimensional representation is
\be
\begin{array}{c}
\mathbf{Q}(u): (1-x_{1} z_{1})^{n}\cdots(1-x_{N}
z_{N})^{n} \mapsto \\
\mapsto \mathrm{S} \cdot (1-\,x_{1} z_{1})^{\frac{n}{2}-u}(1-x_{1}
z_{2})^{\frac{n}{2}+u}\cdots(1-x_{N} z_{N})^{\frac{n}{2}-u}(1-x_{N}
q^{-1}z_{1})^{\frac{n}{2}+u} \,
\end{array}
\ee
and follows from (\ref{bfQ2}) and the explicit action for
the renormalized $\mathrm{Q}_2(u|q)$ (\ref{regQ24}).

\section{Discussion}
We have analyzed the set of commuting operators of the closed
homogeneous spin chain, where the quantum states on the sites are
representations of $s\ell_2$ either infinite-dimensional for
generic spin values or finite-dimensional for integer or
half-integer spins. These operators and their spectra contain all
information on the quantum system. The aim of studying the
relations between them, the Baxter relations in particular, is to
obtain this information in a most convenient and explicit form.

Comparing the ordinary transfer matrix $\mathrm{t}(u)$, its generalizations
$\mathrm{t}_n(u), \mathrm{T}_s(u)$ and the Baxter operators
$\mathrm{Q}(u)$ a simple systematics in their construction is observed
resulting in an understanding of their relations. All they are constructed
as traces of products of operators with one factor for each chain site.
The factor operators act on the tensor product of the quantum and the
auxiliary spaces. Performing construction for infinite-dimensional representations
in the quantum space at generic spin $\ell$ we have seen that
in the most general case the factor at site $k$ is the
general Yang-Baxter operator $\R_{k0}$. In the other cases the factor operators are certain
reductions obtained
therefrom by imposing conditions on the representation parameters $v_1, v_2$
referring to the auxiliary space:

\vspace{0.4cm}
\begin{tabular}{ccc}
{\it chain operator} \ \ & {\it site operator}\ \ \  & {\it restriction} \cr
 $\mathrm{T}_s$ & $\R_{k0}$                   & --- \cr
$\mathrm{t} $    & $\mathrm{L}_k \sim \mathbf{R}_{k0}(\ell,\frac{1}{2})$&
$v_2-v_1 = 2$ and  $\Pi_0^1$ \cr
$\mathrm{t}_n$    & $\mathbf{R}_{k0}(\ell,\frac{n}{2})$ &
$v_2-v_1 = n+1 $ and  $\Pi_0^n$ \cr
$\mathrm{Q}_1 $ & $\R^1_{k0}$ & $v_2 = u_2 $\cr
$\mathrm{Q}_2 $ & $\R^2_{k0}$ & $v_1 = u_1$ \cr
$\mathrm{P}$ & $\mathrm{P}_{k0}$  & $ v_1 = u_1$ and $v_2 = u_2 $ \cr
\end{tabular}
\vspace{0.4cm}

Our proofs of factorization and commutativity for different
transfer matrices rely on local three-term relations --
Yang-Baxter relations. We consider the derivation of these
factorizations as one of the main results due to its transparency
and simplicity. An appropriate case of such relations describes
the intertwining of chain site operators. The standard argument
going parallel to the proof of the ordinary transfer matrix
commutativity then leads to the desired relation for the chain
operators. In particular we have demonstrated how to deduce
algebraic properties of Baxter operators only from these relations
without any references to other concepts.

We have presented two ways of deriving the Baxter relations. The
systematics applies in analogy also to the case of integer or
half-integer spin $\ell = \frac{n}{2}$ with finite-dimensional
representation spaces at the sites. Here the site operators are
$\mathbf{R}_{k0}(u|\frac{n}{2}, s)$ in the  case of the general
transfer matrix $\mathbf{T}_s$ the Yang-Baxter operators
restricted to the irreducible subspace by means of projector
$\Pi^n_k$ at $u_2-u_1 = n+1$. Additional restrictions on parameter
$v_1, v_2$ lead to the other reductions:

\vspace{0.4cm}
\begin{tabular}{ccc}
{\it chain operator} \ \ & {\it site operator}\ \ \  & {\it additional restriction}\cr
 $\mathbf{T}_s$ & $\mathbf{R}_{k0}(\frac{n}{2}|s)$                  & --- \cr
$\mathbf{Q}_1$ or $\mathbf{P}$ & $\mathbf{R}^1_{k0}$ & $v_2 = u_2 $   \cr
$\mathbf{Q}_2$ or $\mathbf{Q}$ & $\mathbf{R}^2_{k0}$ & $v_1 = u_1$ \cr
$\mathbf{S}$ & $\mathbb{S}_{k0}\Pi^n_k$  & $ v_1 = u_1$ and $v_2 = u_2 $ \cr
\end{tabular}
\vspace{0.4cm}

The presented proofs of factorization and commutativity of transfer matrices
for finite-dimensional representations are completely parallel to the
corresponding proofs of the infinite-dimensional case.
Our analysis clarifies the relation between the cases of generic spins
and half-integer or integer spins.
We have seen that the naive attempt to substitute integer or half-integer
values for the spin parameter
$\ell=\frac{n}{2}$ in the general expressions for the regularized transfer
matrices meets difficulties in cases related to the  operator $\mathrm{Q}_2$
just because this operator maps beyond the finite-dimensional quantum space,
whereas the general transfer matrix $\mathrm{T}_s$
and the operator $\mathrm{Q}_1$ do not map beyond the finite-dimensional
quantum space.
Careful calculations of limits $\ell=\frac{n}{2}$ in the general formulae produces
the transfer matrix $\mathrm{S}$ constructed from local operators $\mathbb{S}_{k0}$
being the
nontrivial analogue of the permutation $\mathrm{P}_{k0}$ appearing in the
finite-dimensional case. The operator $\mathrm{S}$ is
a reduction of Baxter operator $\mathrm{Q}_1$ and it
improves the action of the  operator $\mathrm{Q}_2$. Due to the
special operator $\mathrm{S}$
appearing naturally in our scheme we obtain the desired set of Baxter operators
acting in finite-dimensional quantum space.

Moreover, besides presenting general formulae for
Baxter operators and proving their algebraic properties
we  present explicit compact expressions for their action on polynomials.
Such expressions in the case of finite-dimensional quantum space
are particularly simple
and appropriate for practical calculations.

Despite of similarities in our considerations of
infinite-dimensional and finite-dimensional cases there are
significant differences between them. In the infinite-dimensional
case the  Baxter operators construction is $s\ell_2$ symmetric
whereas in finite-dimensional case we had to introduce a symmetry
breaking regularization of divergent traces by means of
q-regularization. We consider this as considerable shortcoming of
the presented construction in the finite-dimensional case.
Actually we have reasons to expect that it should be possible to
construct Baxter operators also in the case of integer or half-integer spin
preserving the symmetry of the model.

\section*{Acknowledgement}

We thank V.Tarasov, A.Manashov and G.Korchemsky for discussions
and critical remarks.

This work has been supported by Deutsche Forschungsgemeinschaft
(KI 623/8-1) One of us is grateful to Leipzig University and DAAD
for support. The work of D.C. is supported by the Chebyshev
Laboratory (Department of Mathematics and Mechanics,
St.-Petersburg State University) under RF government grant
11.G34.31.0026. The work of S.D is supported by the RFFI grants
11-01-00570-а, 11-01-12037, 09-01-12150.
The work of D.K. is supported by Armenian grant 11-1c028.

\appendix
\renewcommand{\theequation}{\Alph{section}.\arabic{equation}}
\setcounter{table}{0}
\renewcommand{\thetable}{\Alph{table}}

\section*{Appendices}

\section{Factorization and commutativity}

\setcounter{equation}{0}

The general transfer matrix constructed from operators
$\R_{k0}(u_{1},u_2|v_1,v_{2})$ factorizes into the product of two
transfer matrices constructed from operators
$\R^{2}_{k0}(u_{1},u_2|v_{2})$ and
$\R^1_{k0^{\prime}}(u_1|v_1,u_{2})$. This factorization for global
objects follows in a clear and direct way from the local relations
for their building blocks.

Let us accomplish analogous steps in order to obtain the second
factorization. We rewrite~(\ref{R1RR}) as follows
$$
\R^1_{00^{\prime}}(v_1|w_1,w_2)
\R_{k0^{\prime}}(u_{1},u_2|w_1,w_2) \R_{k0}(u_{1},u_2|v_1,v_{2}) =
$$
\be \label{rf1'}
=\R_{k0}(u_{1},u_2|v_1,w_{2})\R_{k0^{\prime}}(u_1,u_{2}|w_1,v_{2})
\R^1_{00^{\prime}}(v_1|w_1,w_2) \ee and then specifying parameters
as $w_1=u_1$ and $w_2=u_2$ we arrive at the intertwining relation
$$
\R^1_{00^{\prime}}(v_1|u_1,u_2)\cdot
\mathrm{P}_{k0^{\prime}}\cdot\R_{k0}(u_{1},u_2|v_1,v_{2})
=\R^1_{k0}(u_{1}|v_1,u_{2})\cdot\R^2_{k0^{\prime}}(u_1,u_{2}|v_{2})
\R^1_{00^{\prime}}(v_1|u_1,u_2)
$$
which leads to the relation for the transfer matrices
\begin{equation}\label{factorII}
\tr_{0^{\prime}}
\left[\mathrm{P}_{10^{\prime}}\cdots\mathrm{P}_{N0^{\prime}}\right]\cdot
\tr_{0} \left[\R_{10}(u_{1},u_2|v_1,v_{2})\cdots
\R_{N0}(u_{1},u_2|v_1,v_{2})\right] =
\end{equation}
$$
= \tr_{0} \left[\R^{1}_{10}(u_{1}|v_1,u_{2})\cdots
\R^{1}_{N0}(u_{1}|v_1,u_{2})\right]\cdot \tr_{0^{\prime}} \left[
\R^2_{10^{\prime}}(u_1,u_{2}|v_{2})\cdots
\R^2_{N0^{\prime}}(u_1,u_{2}|v_{2})\right].
$$
The direct consequence of the two factorizations (\ref{factorI})
and (\ref{factorII}) is the commutativity of the transfer matrices
constructed from $\R^1$ and $\R^2$. However it is more instructive
to derive commutativity from local intertwining relations. Below
for completeness we list  the necessary relations.

The Yang-Baxter equation~(\ref{RRR}) in the form \be
\label{PRPRPR} \R_{00^{\prime}}(v_{1},v_2|w_{1},w_2)
\R_{k0^{\prime}}(u_{1},u_2|w_1,w_2) \R_{k0}(u_1,u_2|v_{1},v_2)=
$$
$$
=\R_{k0}(u_{1},u_{2}|v_1,v_2)
\R_{k0^{\prime}}(u_1,u_{2}|w_1,w_{2})
\R_{00^{\prime}}(v_{1},v_2|w_{1},w_2) \ee leads to the
commutativity of the general transfer matrices constructed from
$\R$-operators. Then specifying parameters in (\ref{PRPRPR}) we
get the following three relations. From the first one ($v_1=u_1 \
, \ w_2=u_2$)
$$
\R_{00^{\prime}}(u_{1},v_2|w_{1},u_2)
\R^{1}_{k0^{\prime}}(u_{1}|w_1,u_2) \R^{2}_{k0}(u_1,u_2|v_2)=
$$
\be \label{RR1R2} =\R^{2}_{k0}(u_{1},u_{2}|v_2)
\R^{1}_{k0^{\prime}}(u_1|w_1,u_{2})
\R_{00^{\prime}}(u_{1},v_2|w_{1},u_2) \ee we obtain immediately
the commutativity of the transfer matrices constructed from $\R^1$
and $\R^2$. The second relation ($u_2=v_2=w_2$)
$$
\R^{1}_{00^{\prime}}(v_{1}|w_{1},u_2)
\R^{1}_{k0^{\prime}}(u_{1}|w_1,u_2) \R^{1}_{k0}(u_1|v_1,u_2)=
$$
\be \label{R1R1R1} =\R^{1}_{k0}(u_{1}|v_{1},u_2)
\R^{1}_{k0^{\prime}}(u_1|w_1,u_{2})
\R^{1}_{00^{\prime}}(v_{1}|w_{1},u_2) \ee leads to commutativity
of the transfer matrices constructed from $\R^1$, and the third
one ($u_1=v_1=w_1$)
$$
\R^{2}_{00^{\prime}}(u_{1},v_2|w_{2})
\R^{2}_{k0^{\prime}}(u_{1},u_2|w_2) \R^{2}_{k0}(u_1,u_2|v_2) =
$$
\be \label{R2R2R2} =\R^{2}_{k0}(u_{1},u_2|v_2)
\R^{2}_{k0^{\prime}}(u_1,u_2|w_{2})
\R^{2}_{00^{\prime}}(u_{1},v_2|w_2) \ee implies commutativity of
the transfer matrices constructed from $\R^2$.

\section{Traces}

\setcounter{equation}{0}

Here we consider calculations of the traces in two basic examples.

\medskip

{\bf Trace in operator $Q_2$}

\medskip

Let us consider the operator of the following general form \be
\label{A} \mathbf{A} =
\mathrm{P}_{10}\,\mathrm{A}_{1}(z_1,\partial_1|z_0)\cdot\mathrm{P}_{20}\,
\mathrm{A}_{2}(z_2,\partial_2|z_0)\, \cdots\,
\mathrm{P}_{N0}\,\mathrm{A}_{N}(z_N,\partial_N|z_0)\ , \ee which
acts in the space
$\mathbb{C}[z_0]\otimes\mathbb{C}[z_1]\otimes\ldots\otimes\mathbb{C}[z_N]$.
The operators $\mathrm{A}_{k}$ acting in the space
$\mathbb{C}[z_0]\otimes\mathbb{C}[z_k]$ are arbitrary functions of
the specified arguments. We are going to compute trace of the
operator $\mathbf{A}$ over the space $\mathbb{C}[z_0]$. At first
step we move permutation operators
$\mathrm{P}_{20}\,\cdots\,\mathrm{P}_{N0}$ to the left and obtain
shift operator $\mathrm{P} = \mathrm{P}_{12}\, \cdots\,
\mathrm{P}_{1N}$
$$
\mathbf{A} =
\mathrm{P}\cdot\mathrm{P}_{10}\,\mathrm{A}_{1}(z_1,\partial_1|z_2)\cdot
\mathrm{A}_{2}(z_2,\partial_2|z_3)\,\cdots\,
\mathrm{A}_{N}(z_N,\partial_N|z_0)
$$
and then we act on $z_0^n$
$$
\left[\mathbf{A}\, z_0^n\right] = \mathrm{P}\cdot z_1^n
\mathrm{A}_{1}(z_0,\partial_0|z_2)\cdot
\mathrm{A}_{2}(z_2,\partial_2|z_3)\,\cdots\,
\mathrm{A}_{N}(z_N,\partial_N|z_1) \, .
$$
The trace of operator is equal to the sum of diagonal matrix
elements
\begin{equation}\label{traceV}
\tr_{0} \mathbf{A} = \sum_{n=0}^{\infty} \frac{1}{n!}
\partial_0^n\left.\left[\mathbf{A}\, z_0^n\right]\right|_{z_0=0} \, .
\end{equation}
Applying this general formula we finally obtain
$$
\tr_{_0}\mathbf{A} = \mathrm{P}\cdot \sum_{n=0}^{\infty}
\frac{z_1^n}{n!} \,
\partial_0^n \left.\mathrm{A}_{1}(z_0,\partial_0|z_2)\cdot
\mathrm{A}_{2}(z_2,\partial_2|z_3)\cdots
\mathrm{A}_{N}(z_N,\partial_N|z_1)\right|_{z_0 =0} =
$$
\begin{equation}\label{traceV1}
= \mathrm{P}\cdot\left.\mathrm{A}_{1}(z_1,\partial_1|z_2)\cdot
\mathrm{A}_{2}(z_2,\partial_2|z_3)\cdots
\mathrm{A}_{N}(z_N,\partial_N|z_0)\right|_{z_0\to z_1}
\end{equation}

This explicit formula indicates that the trace over infinite
dimensional space of operator (\ref{A}) converge without any additional
regularization.
\medskip

{\bf Trace in operator $Q_1$}

\medskip

Consider the formula
$$
\sum_{n=0}^\infty\frac1{n!}\dd^n\,(B+Az)^n\,\Phi(z)\biggl|_{z=0} =
\sum_{n=0}^\infty
\dd^n\,\left[\sum_{k=0}^n\frac{B^{k}A^{n-k}}{(n-k)!k!}
\sum_{j=0}^\infty\frac1{j!}\Phi^{(j)}(0)\,z^{n-k+j}\right]\biggl|_{z=0}
=
$$
$$
=\sum_{n=0}^\infty\sum_{k=0}^n\frac{n!B^kA^{n-k}}{(n-k)!(k!)^2}
\,\Phi^{(k)}(0) = \sum_{k=0}^\infty\frac{B^k\,
\Phi^{(k)}(0)}{k!}\sum_ {n=k}^\infty\frac{n!A^{n-k}}{(n-k)!k!} =
\sum_{k=0}^\infty \frac{B^k\Phi^{(k)}(0)}{k!(1-A)^{k+1}}\,.
$$
Here at the first step the expression in square bracket has been
expanded in power series, then the $n$-th derivative has been
taken and the condition $z=0$ has been imposed, which imposes $j =
k$. After that the order of summations has been changed and the
inner sum has been summed up. The last expression is just Taylor
series of $\frac{1}{1-A}\Phi\left(\frac{B}{1-A}\right)$.

Now we are going to obtain the formula (\ref{TraceExpl}).
The proof is based on the local intertwining relation
$$
\mathrm{P}_{k0} \lambda^{z_{0k}\dd_0} \cdot \mathrm{P}_{k0'}(1-\textstyle{\frac{1}{\lambda}})^{z_{k0'}\dd_k}\cdot
\mathrm{P}_{00'}\lambda^{z_{00'}\dd_{0'}}  =
\mathrm{P}_{00'}\lambda^{z_{00'}\dd_{0'}} \cdot \mathrm{P}_{k0'} \cdot
\mathrm{P}_{k0}\lambda^{z_{0k}\dd_0}(1-\textstyle{\frac{1}{\lambda}})^{z_{k0}\dd_k}
$$
which can be checked straightforwardly. It produces factorization relation for
corresponding transfer matrices
\be \label{factApp}
\mathrm{A} \cdot \mathrm{B}
= \mathrm{P}\cdot \mathrm{C}
\ee
where
\bea
\mathrm{A}&\equiv& \tr_{\mathbb{V}_{0}} \mathrm{P}_{10} \lambda_1^{z_{01}\dd_0}\cdots\mathrm{P}_{N0} \lambda_N^{z_{0N}\dd_0} \\
\mathrm{B} &\equiv & \tr_{\mathbb{V}_{0'}} \mathrm{P}_{10'}
(1-\textstyle{\frac{1}{\lambda_1}})^{z_{10'}\dd_{1}}\cdots
\mathrm{P}_{N0'} (1-\textstyle{\frac{1}{\lambda_N}})^{z_{N0'}\dd_{N}} \\
\mathrm{C} &\equiv &  \tr_{\mathbb{V}_{0}} \mathrm{P}_{10}\lambda_1^{z_{01}\dd_0}(1-\textstyle{\frac{1}{\lambda_1}})^{z_{10}\dd_1}
\cdots\mathrm{P}_{N0}\lambda_N^{z_{0N}\dd_0}(1-\textstyle{\frac{1}{\lambda_N}})^{z_{N0}\dd_N}
\eea
Applying (\ref{tr21}) with
$\lambda\to1-\frac{1}{\lambda}$
we obtain immediately
$$
\mathrm{B}\cdot\Psi(\vec{z}\,) = \mathrm{P} \,\Psi(\Lambda'\vec{z}\,)\,.
$$
Then we proceed to the calculation of the trace in the definition of $\mathrm{C}$.
Its building blocks act on the function as follows
$$
\mathrm{P}_{k0}\lambda^{z_{0k}\dd_0}(1-\textstyle{\frac{1}{\lambda}})^{z_{k0}\dd_k} \, \Phi(z_k , z_0)
= \Phi(z_k , \lambda z_k + \bar{\lambda} z_0)\,.
$$
Consequently being applied to the function $\Psi(\vec{z}\,)$ operator
$\mathrm{C}$ does not change its arguments and only produces
overall factor which we calculate using the formula (\ref{sSum})
$$
\mathrm{C} \cdot  \Psi(\vec{z}\,) = \frac{1}{ 1 - \bar{\lambda}_1 \cdots \bar{\lambda}_N } \cdot \Psi(\vec{z}\,)\,.
$$
Finally we act by both sides of (\ref{factApp}) on the function $\Psi(\vec{z}\,)$
and obtain
$$
\mathrm{A}\cdot\mathrm{P}\,\Psi(\Lambda'\vec{z}\,) =
\frac{1}{ 1 - \bar{\lambda}_1 \cdots \bar{\lambda}_N } \cdot \mathrm{P} \, \Psi(\vec{z}\,) \,
$$
which can be casted in the form (\ref{TraceExpl}) by means of a linear transformation
of $\vec{z}$.

\section{q-Regularization}

\setcounter{equation}{0}

In this Appendix we present the necessary modifications of the
formulae of  sections~\ref{Qbyhand} and~\ref{global}  after
introduction of the $q$-regularization. We use the following
regularization
$$
\tr_{_0}\mathbf{B} \longrightarrow \tr_{_0} q^{z_0 \dd_0} \mathbf{B} \, , \qquad |q|<1
$$
Since the  boundary conditions change to quasiperiodic ones the
shift operator takes the form
 \be \label{qTrP} \tr_{0}
\left[\, q^{z_0 \dd_0} \cdot
\mathrm{P}_{10}\cdots\mathrm{P}_{N0}\right] =
\mathrm{P}_{12}\mathrm{P}_{13}\cdots\mathrm{P}_{1N} \cdot \tr_{0}
\left[\, q^{z_0 \dd_0} \mathrm{P}_{10} \right] = \mathrm{P} \cdot
q^{z_1 \dd_1} \,. \ee

The introduction of the regularization in the factorization relation
(\ref{factorI}) leads to
$$
\mathrm{P}\, q^{z_1 \dd_1} \cdot \tr_{0} \left[ \, q^{z_0 \dd_0}
\, \R_{10}(u_{1},u_2|v_1,v_{2})\cdots
\R_{N0}(u_{1},u_2|v_1,v_{2})\right] =
$$
\begin{equation} \label{regfactorI}
= \tr_{0} \left[ \, q^{z_0 \dd_0}
\R^{2}_{10}(u_{1},u_2|v_{2})\cdots
\R^{2}_{N0}(u_{1},u_2|v_{2})\right]\cdot \tr_{0^{\prime}} \left[
\, q^{z_{0'} \dd_{0'}}\R^1_{10^{\prime}}(u_1|v_1,u_{2})\cdots
\R^1_{N0^{\prime}}(u_1|v_1,u_{2})\right]
\end{equation} whereas
the second factorization (\ref{factorII}) modifies in the
following manner
$$
\mathrm{P}\, q^{z_1 \dd_1} \cdot
\tr_{0} \left[ \, q^{z_0 \dd_0} \,
\R_{10}(u_{1},u_2|v_1,v_{2})\cdots
\R_{N0}(u_{1},u_2|v_1,v_{2})\right] =
$$
\begin{equation} \label{regfactorII}
= \tr_{0}\left[\, q^{z_0 \dd_0} \R^{1}_{10}(u_{1}|v_1,u_{2})\cdots
\R^{1}_{N0}(u_{1}|v_1,u_{2})\right]\cdot\tr_{0^{\prime}} \left[ \,
q^{z_{0'} \dd_{0'}}\R^2_{10^{\prime}}(u_1,u_{2}|v_{2})\cdots
\R^2_{N0^{\prime}}(u_1,u_{2}|v_{2})\right]
\end{equation}

The regularized general transfer matrix has the form \be
\label{regT} \mathrm{T}_{s}(u|q) = \tr_0 \left[ \, q^{z_0 \dd_0}
\, \R_{10}(u|\ell, s) \R_{20} (u|\ell, s) \cdots \R_{N0} (u|\ell,
s) \right] \,. \ee Specifying the parameters in the transfer
matrix one obtains the regularized Baxter $\mathrm{Q}$-operators
$$
\mathrm{Q}_1(u-v_1|q) = \tr_{0} \left[ \, q^{z_0 \dd_0} \,
\R^{1}_{10}(u_1|v_1,u_{2})\cdots
\R^{1}_{N0}(u_1|v_1,u_{2})\right]\,,
$$
\be \label{regQ} \mathrm{Q}_2(u-v_2|q) = \tr_{0} \left[ \, q^{z_0
\dd_0} \, \R^{2}_{10}(u_{1},u_2|v_{2})\cdots
\R^{2}_{N0}(u_{1},u_2|v_{2})\right]\,.\ee Regularized commutation
relations have the form \be \label{regT->Q1Q2} \mathrm{P}\, q^{z_1
\dd_1}\cdot\mathrm{T}_{s} (u|q) = \mathrm{Q}_2 (u - s|q) \,
\mathrm{Q}_1 (u + s + 1|q) = \mathrm{Q}_1 (u + s + 1|q) \,
\mathrm{Q}_2 (u - s|q) \ee and the set of commutation relations is
\be \label{commutRegPQi} [ \, \mathrm{P}\, q^{z_1 \dd_1} \, , \,
\mathrm{Q}_{1}(u|q) \, ] = [ \, \mathrm{P}\, q^{z_1 \dd_1} \, , \,
\mathrm{Q}_{2}(u|q) \, ] = 0 \,, \ee \be \label{commutRegQiQk} [
\, \mathrm{Q}_1 (u|q) , \mathrm{Q}_1 (v|q) \, ] = [ \,
\mathrm{Q}_2 (u|q) , \mathrm{Q}_2 (v|q) \, ] = [ \, \mathrm{Q}_1
(u|q) , \mathrm{Q}_2 (v|q) \, ] = 0 \,, \ee \be
\label{commutRegTQi} [ \, \mathrm{T}_{s}(u|q) , \mathrm{Q}_1 (v|q)
\, ] = [ \, \mathrm{T}_{s}(u|q) , \mathrm{Q}_2 (v|q) \, ] = 0 \,.
\ee
Thus all the formulae of section \ref{global} for different
transfer matrices preserve their form provided they have been
substituted by their regularized version. It is important that the
regularization parameter is the same in all the formulae.

Let us turn to the Baxter equation. In the section~\ref{Qbyhand}
we did not use any trace at all so that all modifications are
exterior and reduce to simple redefinitions of the main
objects. First of all the ordinary transfer matrix $\mathrm{t}(u)$
is changed
\be  \mathrm{t}(u|q) = \tr
\begin{pmatrix} q & 0 \\ 0 & 1 \end{pmatrix}
\mathrm{L}_1(u)\mathrm{L}_2(u)\cdots \mathrm{L}_N(u)\,. \ee
Further the definition of the $\mathrm{Q}_2$-operator is changed
\begin{equation}\label{regQ_2}
\mathrm{Q}_2(u|q) = \mathrm{P}\,
q^{z_1\partial_1}\cdot\left.\mathrm{R}_{12}^{2}(u)\,
\mathrm{R}_{23}^{2}(u)\, \cdots\, \mathrm{R}_{N-1,N}^{2}(u)
\mathrm{R}_{N0}^{2}(u)\right|_{z_0\to \frac{z_1}{q}}\
\end{equation}
because we want to keep the consistency with representation of
this operator as transfer matrix constructed from the
$\R^2$-operators.
The regularized analog of the trace formula in
 Appendix A has the following form
$$
\tr_{_0} \left[ \, q^{z_0 \dd_0} \cdot
\mathrm{P}_{10}\,\mathrm{A}_{1}(z_1,\partial_1|z_0)\cdot\mathrm{P}_{20}\,
\mathrm{A}_{2}(z_2,\partial_2|z_0)\, \cdots\,
\mathrm{P}_{N0}\,\mathrm{A}_{N}(z_N,\partial_N|z_0)\, \right] =
$$
\be \label{qTrace} = \mathrm{P}\, q^{z_1\partial_1}\cdot
\left.\mathrm{A}_{1}(z_1,\partial_1|z_2)\cdot
\mathrm{A}_{2}(z_2,\partial_2|z_3)\cdots
\mathrm{A}_{N}(z_N,\partial_N|z_0)\right|_{z_0= \frac{z_1}{q}}\,
\ee which fixes the needed modifications in definition of the
$\mathrm{Q}$-operator. These changes lead to corresponding
modification of the Baxter equation
$$ \mathrm{t}(u|q) \,
\mathrm{Q}_2(u|q) = q\cdot \mathrm{Q}_2(u+1|q)+ (u_1 u_2)^N
\cdot\mathrm{Q}_2(u-1|q)\,
$$
due to changes in the main formula (\ref{mainf})
$$
\begin{pmatrix} 1 & 0 \\ -\frac{z_1}{q} & 1
\end{pmatrix}\cdot q^{z_1 \dd_1}\,\mathrm{R}_{12}^{2}(u)
\mathrm{R}_{23}^{2}(u) \cdots \mathrm{R}_{N0}^{2}(u)\cdot
\begin{pmatrix} q & 0 \\ 0 & 1 \end{pmatrix}
\mathrm{L}_1(u)\mathrm{L}_2(u)\cdots\mathrm{L}_N(u)\cdot\begin{pmatrix}
1 & 0 \\ z_0 & 1
\end{pmatrix}
=
$$
$$
=
q^{z_1 \dd_1}\,\begin{pmatrix} q & 0 \\ 0 & 1 \end{pmatrix}\left(%
\begin{array}{cc}
\mathrm{R}^{2}_{12}(u+1) & -\mathrm{R}^{2}_{12}(u)\partial_{1} \\
 0 & u_1\,u_2\,\mathrm{R}^{2}_{12}(u-1) \\
\end{array}%
\right)\cdots\left(%
\begin{array}{cc}
\mathrm{R}^{2}_{N0}(u+1) & -\mathrm{R}^{2}_{N0}(u)\partial_{N} \\
 0 & u_1\,u_2\, \mathrm{R}^{2}_{N0}(u-1) \\
\end{array}%
\right)
$$
and the modification in the formula for the action on the generating
function
\begin{equation}
\mathrm{Q} (u|q): (1-x_{1} z_{1})^{2\ell}\cdots(1-x_{N}
z_{N})^{2\ell} \mapsto \label{regQ24}
\end{equation}
$$
\mapsto (1-q \,x_{1} z_{N})^{\ell-u}(1-x_{1}
z_{1})^{\ell+u}\cdots(1-x_{N} z_{N-1})^{\ell-u}(1-x_{N}
z_{N})^{\ell+u} \, .
$$
Let us consider the modifications in the uniform approach to the
derivation of Baxter equation for both $\mathrm{Q}$-operators. It
turns out  that the Baxter equations are different for $q$-regularized
operators $\mathrm{Q}_1$ and $\mathrm{Q}_2$. The transfer matrix which
is constructed as the trace over finite dimensional space
$\mathbb{V}_n$ has the form
\be \label{regt_n} \mathrm{t}_{n}(u|q)
= \tr \, q^{z_0 \dd_0} \, \mathbf{R}_{10}(u|\ell ,
{\textstyle\frac{n}{2}} \, ) \, \mathbf{R}_{20}(u|\ell ,
{\textstyle\frac{n}{2}}\, ) \, \ldots \, \mathbf{R}_{N0}(u|\ell ,
{\textstyle\frac{n}{2}} \, )\,. \ee
Because
$$
\mathcal{D} \, q^{z_0 \dd_0} = q^{n+1} \cdot q^{z_0 \dd_0} \mathcal{D}\,,
$$
we have the modification \footnote{Now we change normalizations of operators $\mathrm{R}$ and $\mathrm{R}^1$ as
in the subsetion \ref{det-formula} in order to deal with Baxter relations.} 

$$
\mathrm{T}_{\frac{n}{2}}(u) = \mathrm{t}_{n}(u)+ q^{n+1} \cdot
\mathrm{T}_{-\frac{n}{2} - 1}(u) \,.
$$
$$
\tr \mathcal{D} \, q^{z_0 \dd_0} \, \R_{10}(u|\ell ,
{\textstyle\frac{n}{2}} \,) \, \R_{20}(u|\ell ,
{\textstyle\frac{n}{2}} \,) \, \ldots \, \R_{N0}(u|\ell ,
 {\textstyle\frac{n}{2}} \,) \, \mathcal{D}^{-1} = q^{n+1} \cdot
 \mathrm{T}_{-\frac{n}{2} - 1 }(u|q)
 $$
and the determinant relation takes the following modified form
$$
\mathrm{P}\, q^{z_1\partial_1}\cdot\mathrm{t}_{n}(u|q) =
\left|\begin{array}{cc}
\mathrm{Q}_1(u+ \frac{n}{2} +1|q) & q^{\beta} \cdot \mathrm{Q}_2(u+ \frac{n}{2} +1|q) \\
q^{\alpha} \cdot \mathrm{Q}_1(u - \frac{n}{2}|q)& \mathrm{Q}_2(u -
\frac{n}{2}|q)\end{array}\right| \, , \quad \text{where} \quad n + 1
= \alpha + \beta\,.
$$
The general bilinear relation involving $\mathrm{Q}_1$ is derived
starting from the following determinant
$$
\left|\begin{array}{ccc}
\mathrm{Q}_1(a|q) & q^{n+1} \mathrm{Q}_2(a|q) & \mathrm{Q}_1(a|q)\\
\mathrm{Q}_1(b|q) & \mathrm{Q}_2(b|q) & \mathrm{Q}_1(b|q)\\
q^{m+1} \mathrm{Q}_1(c|q) & \mathrm{Q}_2(c|q) & q^{m+1}
\mathrm{Q}_1(c|q)
\end{array}\right| = 0
$$
Specifying  parameters
$$
a = u + \frac{n}{2} + 1 \ ; \ b = u - \frac{n}{2}\ ; \ c = u
-m-\frac{n}{2} - 1 \, ,
$$
we arrive at the relation
$$ \mathrm{t}_{m}\left(u-1-{\textstyle\frac{n+m}{2}}| q\right)\cdot
\mathrm{Q}_1\left(u+1+{\textstyle\frac{n}{2}}|q\right)
-\mathrm{t}_{n+m+1}\left(u-{\textstyle\frac{m+1}{2}}|q\right)\cdot
\mathrm{Q}_1\left(u-{\textstyle\frac{n}{2}}|q\right) +
$$
$$
+ \, q^{m+1}\cdot\mathrm{t}_{n}\left(u |q\right)\cdot
\mathrm{Q}_1\left(u-1-m-{\textstyle\frac{n}{2}}|q\right) = 0 \,.
$$
In a special case $n = m =0$ one obtains the Baxter equation for
the operator $\mathrm{Q}_1(u|q)$
$$
\mathrm{t}(u|q)\mathrm{Q}_1(u|q) =\mathrm{Q}_1(u+1|q)+  q\cdot(u_1
u_2)^N \cdot\mathrm{Q}_1(u-1|q)\,.
$$
For completeness we derive
the Baxter equation for the second operator $\mathrm{Q}_2$. All the
difference is in the starting point
$$
\left|\begin{array}{ccc}
\mathrm{Q}_1(a|q) & q^{n+1} \mathrm{Q}_2(a|q) & q^{n+1} \mathrm{Q}_2(a|q)\\
\mathrm{Q}_1(b|q) & \mathrm{Q}_2(b|q) & \mathrm{Q}_2(b|q)\\
q^{m+1} \mathrm{Q}_1(c|q) & \mathrm{Q}_2(c|q) & \mathrm{Q}_2(c|q)
\end{array}\right| = 0\,.
$$
Next we specify parameters in the same way and arrive at the  relation
$$ q^{n+1}\cdot\mathrm{t}_{m}\left(u-1-{\textstyle\frac{n+m}{2}}| q\right)\cdot
\mathrm{Q}_2\left(u+1+{\textstyle\frac{n}{2}}|q\right)
-\mathrm{t}_{n+m+1}\left(u-{\textstyle\frac{m+1}{2}}|q\right)\cdot
\mathrm{Q}_2\left(u-{\textstyle\frac{n}{2}}|q\right) +
$$
$$
+ \, \mathrm{t}_{n}\left(u |q\right)\cdot
\mathrm{Q}_2\left(u-1-m-{\textstyle\frac{n}{2}}|q\right) = 0\, ,
$$
and at the special point $n = m = 0$ one obtains the Baxter
equation~(\ref{qBaxter2}) for the second operator.

\end{document}